\documentclass[letterpaper, 12pt]{iopart}

\expandafter\let\csname equation*\endcsname\relax
\expandafter\let\csname endequation*\endcsname\relax
\usepackage{amsthm, amssymb}
\usepackage[intlimits]{amsmath}
\usepackage{bm}

\usepackage{graphicx}

\usepackage{etex}
\usepackage{mathrsfs}
\usepackage{microtype}
\usepackage{booktabs}
\usepackage{tabularx}
\usepackage{color}
\usepackage{pifont}
\usepackage[square, comma, sort&compress, numbers]{natbib}

\usepackage[sc, font=footnotesize]{caption}

\usepackage{subfigure}

\usepackage{fancyvrb}

\usepackage[pdfauthor={P.H. Trinh, M.J. Ward},pdftitle={Dynamics of localized spot patterns on the sphere}]{hyperref}

\usepackage{thmtools, thm-restate}

\newtheorem{result}{Principal Result}
\newtheorem{lemma}{Lemma}

\newcommand{\de}{\mathrm{d}}
\newcommand{\dd}[2]{\frac{\de#1}{\de#2}}
\newcommand{\pd}[2]{\frac{\partial#1}{\partial#2}}

\newcommand{\Oh}{\mathcal{O}}

\newcommand*\out{\mathrm{out}}

\newcommand*\unif{\mathrm{unif}}
\newcommand*\uin{u_\mathrm{in}}
\newcommand*\vin{v_\mathrm{in}}
\newcommand*\dtheta{\dot{\theta}}
\newcommand*\dphi{\dot{\phi}}

\newcommand*{\UVb}{\begin{pmatrix}U_{j1} \\ V_{j1}\end{pmatrix}}
\newcommand*{\UVeq}{\begin{pmatrix} U_{j1}^{\text{e}} \\ V_{j,1}^{\text{e}}\end{pmatrix}}
\newcommand*{\UVdyn}{\begin{pmatrix} U_{j1}^{\text{d}} \\ V_{j,1}^{\text{d}}\end{pmatrix}}

\newcommand*{\Ueq}{\bm{U}_{j1}^{\text{e}}}
\newcommand*{\Udyn}{\bm{U}_{j1}^{\text{d}}}

\newcommand*{\B}{\mathcal{B}}
\newcommand*{\K}{\mathcal{K}}
\newcommand*{\RC}{\mathcal{R}}
\newcommand*{\T}{\mathcal{T}}
\newcommand*{\E}{\mathrm{E}}
\newcommand*{\Z}{\mathcal{Z}}

\newcommand*{\I}{\bm{\mathrm{I}}}

\newcommand*{\G}{\mathcal{G}}
\newcommand*{\Lsum}{\mathcal{L}}

\DeclareMathOperator{\dell}{\Delta}
\newcommand*{\enn}{\mathfrak{N}}
\newcommand*{\tee}{\mathfrak{T}}
\newcommand*{\Lop}{\mathfrak{L}}

\newcommand*{\N}{\mathcal{N}}
\newcommand*{\M}{\mathcal{M}}

\newcommand*{\Fu}{F}
\newcommand*{\Fv}{H}
\newcommand*{\R}{{\mathbb{R}}}

\newcommand{\eps}{{\displaystyle \varepsilon}}

\newcommand{\bsub}{\begin{subequations}}
\newcommand{\esub}{\end{subequations}$\!$}

\begin{document}

\title[Dynamics of localized spot patterns on the sphere]
{The dynamics of localized spot patterns for reaction-diffusion systems
on the sphere}

\author{Philippe H. Trinh$^1$ and Michael J. Ward$^2$}

\address{$^1$ Oxford Centre for Industrial and Applied Mathematics, Mathematical Institute, University of Oxford, Oxford, Oxfordshire, OX2 6GG \\
$^2$ Department of Mathematics,  University of British Columbia, \\
Vancouver, British Columbia, V6T 1Z2
}
\eads{\mailto{trinh@maths.ox.ac.uk}, \mailto{ward@math.ubc.ca}}
\begin{abstract}
In the singularly perturbed limit corresponding to a large diffusivity ratio between two components in a reaction-diffusion (RD) system, quasi-equilibrium spot patterns are often admitted, producing a solution that concentrates at a discrete set of points in the domain. In this paper, we derive and study the differential algebraic equation (DAE) that characterizes the slow dynamics for such spot patterns for the Brusselator RD model on the surface of a sphere. Asymptotic and numerical solutions are presented for the system governing the spot strengths, and we describe the complex bifurcation structure and demonstrate the occurrence of imperfection sensitivity due to higher order effects.  Localized spot patterns can undergo a fast time instability and we derive the conditions for this phenomena, which depend on the spatial configuration of the spots and the parameters in the system.  In the absence of these instabilities, our numerical solutions of the DAE system for $N = 2$ to $N = 8$ spots suggest a large basin of attraction to a small set of possible steady-state configurations.  We discuss the connections between our results and the study of point vortices on the sphere, as well as the problem of determining a set of elliptic Fekete points, which correspond to globally minimizing the discrete logarithmic energy for $N$ points on the sphere.
\end{abstract}


\section{Introduction}

\noindent We analyze localized spot patterns for a two-component
reaction-diffusion (RD) system on the surface of a sphere. In the
singularly perturbed limit that corresponds to the large diffusivity
ratio, such systems will often permit the formation of spatially
localized spot patterns, These patterns are characterized by one or both solution
components concentrating at certain points in the domain. At
leading-order, the spot patterns are stationary, and in a companion
paper by Rozada \emph{et al.}~\cite{rozada2014}, results for these
quasi-equilibria structures were presented for the prototypical model
of the Brusselator. Over long time scales, however, and for finite diffusivity ratios, the spots will indeed move on the sphere. The main goal of this paper is to derive and analyze
these resultant slow spot dynamics.

We focus our analysis on the dimensionless Brusselator system 
given in terms of the activator $u = u(\bm{x},t)$ and the inhibitor $v
= v(\bm{x},t)$ on the surface of the unit sphere, formulated as
\begin{subequations}\label{full_all}
\begin{equation} \label{full}
\pd{u}{t}  = \eps^2 \dell_S u + \Fu(u, v)\,, \qquad
\tau \pd{v}{t} = \dell_S v +  \Fv(u,v)\,, 
\end{equation}
where the nonlinear kinetics are defined by
\begin{equation}\label{full:kinetics}
\Fu(u,v) \equiv \eps^2 \E - u + fu^2 v\,, \qquad
\Fv(u,v) \equiv \eps^{-2}\left(u - u^2 v\right) \,,
\end{equation}
\end{subequations}
for constants $\E > 0$, $\tau>0$, and $0 < f < 1$. In (\ref{full}),
the surface Laplacian, $\dell_S$, is defined by
\begin{equation} \label{laplace}
\dell_S \equiv \frac{1}{\sin^2 \theta} \pd{^2}{\phi^2} + 
\frac{1}{\sin\theta} \pd{}{\theta} \left[\sin\theta \pd{}{\theta}\right] \,,
\end{equation}
corresponding to the spherical coordinate system $\bm{x} = (x,y,z) =
(\cos\phi \sin \theta, \sin\phi \sin\theta, \cos\theta)^{T}$, for
longitudinal angular coordinate $\phi \in [0, 2\pi)$ and latitudinal
  coordinate $\theta \in (0, \pi)$. The particular scaling of the
  non-dimensionalized system \eqref{full_all} has been primarily
  chosen so that the magnitude of the spot patterns for $u$ is
  $\Oh(1)$ in the limit $\eps \to 0$. In \ref{sec:brus}, we review the
  full details of the scalings leading to \eqref{full_all}, as given in
  \cite{rozada2014}.

\begin{figure}[htb]\centering
\includegraphics[width=0.9\textwidth]{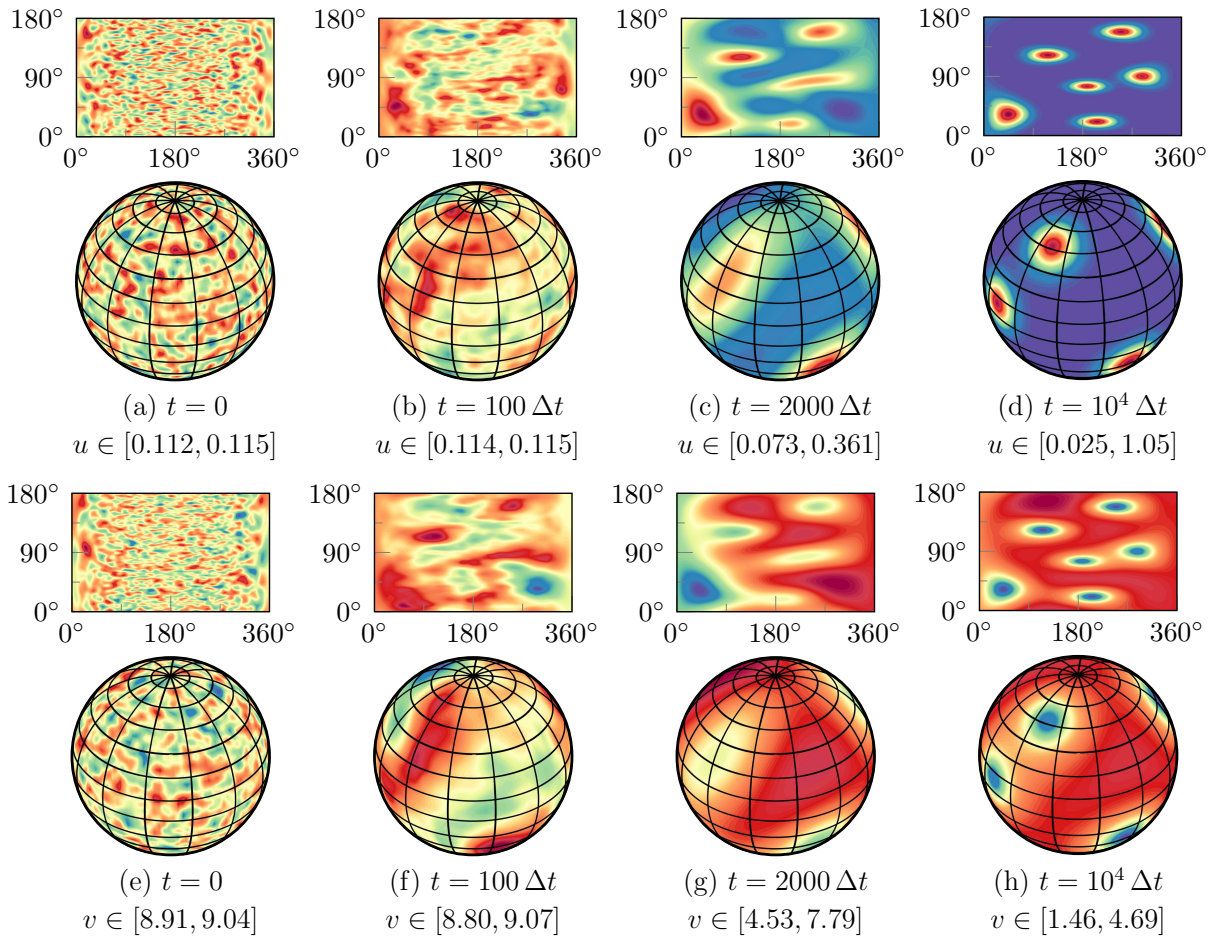}
\caption{Full numerical solutions $u$ in (a) to (d), and $v$ in (e) to (h) of the RD system
  (\ref{full_all}) computed using the closest-point method used in
  \cite{rozada2014} with explicit Euler time integration. The parameter
  values are $f = 0.8$, $\eps = 0.075$, $\tau = 7.8125$,
  $\E = 4$. Time steps were $\Delta t = 0.005$ and $\Delta x = \Delta
  y = 0.08$. Blue denotes small values, yellow middle values, and red large values. The top
  subplots display the patterns in the $(\phi, \theta)$
  plane. \label{fig:evo}}
\end{figure}

For small values of $\eps$, localized spot patterns are readily
observed in full numerical simulations of \eqref{full_all} when using
random initial conditions close to the spatially uniform state
$u_e={\eps^2\E/(1-f)}$ and $v_e={(1-f)/(\E \eps^2)}$. For example,
using one set of parameter values and with a $1\%$ random perturbation
of the uniform state, Fig.~\ref{fig:evo} shows that the intricate
transient dynamics at short times leads to the formation of six
localized spots as time increases. Thus, given that spot-type patterns
can emerge in the singularly perturbed limit, $\eps\to 0$, it is of
interest to asymptotically construct such patterns and then to analyze
their stability and slow dynamics. A central question is to ask
whether, beginning from an $N$-spot pattern, one can asymptotically
derive from (\ref{full_all}) a reduced dynamical system for the time
evolution of the spot centers. From this limiting system, one can then
determine the spatial locations of the centers of the spots that
correspond to linearly stable steady-state patterns on the sphere.

\subsection{Extending from the quasi-equilibrium study of Rozada \emph{et al.}~\cite{rozada2014}}


Our understanding of the quasi-static spot patterns on the sphere
relies upon many results presented in the companion paper by Rozada
\emph{et al.}~\cite{rozada2014}. There, the method of matched
asymptotic expansions was used in the limit $\eps\to 0$ to construct
the quasi-static $N$-spot solution for (\ref{full_all}), with the
spots centered at $\bm{x}_1,\ldots,\bm{x}_N$ on the sphere. In the
outer region, defined at ${\mathcal O}(1)$ distances from the spot
locations, it was shown that the leading-order inhibitor
concentration field, $v$, in (\ref{full_all}) is given in terms of a
sum of Green's functions, where each spot is represented as a Coulomb
singularity of the form $v\sim S_j\log|\bm{x}-\bm{x}_j|$ as
$\bm{x}\to\bm{x}_j$, for $j=1,\ldots,N$. The spot strengths
$S_1,\ldots,S_N$ were found to satisfy a nonlinear algebraic system
involving a Green's matrix, representing interactions between the
spots, and a nonlinear function arising from the local solution near
an individual spot. An important parameter that is introduced was
\begin{equation}
  \nu = \frac{1}{|\log \eps|},
  \end{equation}
which arises during the matching process between the outer solutions,
valid away from the spot centers, and the inner core solutions. 
This gauge function results from the logarithmic singularity
of the Green's function on the sphere.

Moreover, in the companion study, it was shown that from a
  numerical solution of a radially symmetric eigenvalue problem that
  if the spot strength exceeds some threshold, then the
  $j^\textrm{th}$ spot is linearly unstable to a non-radially
  symmetric peanut-shape perturbation near the spot. This linear
  instability was found to be the trigger of a nonlinear spot
  self-replication event, suggesting that this bifurcation is
  subcritical.  In addition, a globally coupled eigenvalue problem
  (GCEP) was formulated that determines the stability properties of an
  $N$-spot pattern to locally radially symmetric perturbations near
  the spots. This GCEP was analyzed in \cite{rozada2014}
  only for special spatial configurations
  $\lbrace{\bm{x}_1,\ldots,\bm{x}_N\rbrace}$ of spots for which they
  have a common strength, \emph{i.e.} $S_c=S_j$ for $j=1,\ldots,N$.

In this paper, we shall build upon the companion
study~\cite{rozada2014} by presenting an asymptotic and numerical
study of the slow spot patterns. We will also provide a more complete
analysis of the quasi-equilibria patterns, particularly noting further
distinguished limits as $\eps \to 0$, and solutions with unequal spot
strengths, which had not been previously uncovered in
\cite{rozada2014}.

Our plan is as follows. First in \S\ref{sec:asym}, we shall derive the
set of equations that governs the slow movement of the spot
locations. We demonstrate that in the absence of any ${\mathcal O}(1)$
time-scale instability, the spots centers will slowly drift on an
asymptotically long time-scale of order $\Oh(\eps^{-2})$. The
governing equations take the form of differential algebraic equation
(DAE) for the time-evolution of the spot locations, which depend on
the current spot strengths. The main technical challenge in deriving
this DAE is due to the higher-order matching between the inner
(near-spot) and outer solutions. In particular, this asymptotic
matching must account for inter-spot interactions, the slow dynamics
of the patterns, and the correction terms that arise due to the
projection the spherical geometry onto the local tangent plane
approximation near the $j^\textrm{th}$ spot.

After having done so, we shall return in \S\ref{sec:qe_instab} to the
study of the quasi-equilibrium solutions and provide a new analysis
that accounts for the distinguished limits that arise when $E$ in
\eqref{full_all} is either $\Oh(1)$ or simultaneously tends to zero
when $\nu \to 0$. We identify a set of patterns of quasi-equilibrium
patterns, not remarked in \cite{rozada2014}, that consists of spots of
mixed strengths, and we demonstrate that such mixed patterns are all
unstable on an $\mathcal{O}(1)$ time-scale. We furthermore extend the
prior study by applying numerical path-following methods to the
nonlinear algebraic system in order to illustrate the bifurcation
structure. Notably, we demonstrate the new result that, in
the regime $\E = \Oh(\sqrt{\nu})$, the bifurcation structure can
exhibit imperfection sensitivity if a certain condition on the spot
locations does not hold.

In \S\ref{sec:dynamics}, we perform numerical simulations of the DAE
system in the parameter regime for which the quasi-equilibrium spot
patterns are linearly stable. By beginning from random initial
configurations for $N = 2$ to $N = 8$ spots, we identify the
steady-state patterns having large basins of attraction. A
particularly difficult configuration to identify is for $N=8$, where
the stable steady-state pattern is a $45^{\circ}$ ``twisted cuboid'':
two parallel rings containing four equally-spaced spots, with the
spots phase shifted by $45^\circ$ between each ring (see
Fig.~\ref{fig:rotate}). In fact, this special 8-spot pattern is an
elliptic Fekete point set. Our main results are summarized in
\S\ref{sec:disc}, and we discuss numerous open problems for future
study.

\subsection{Connections and differences with other works}

There is a wealth of literature on the formation of RD patterns in
both simple and more complicated domains. Let us review how the work
our this paper fits into the wider community.

For the situation where only one of the two solution components is
localized, the spots are said to exhibit \emph{semi-strong
  interactions}. In this semi-strong interaction limit, and in a 1-D
spatial domain, there have been many studies of the dynamics of
localized patterns for specific reaction-diffusion systems; this
includes the Gierer-Meinhardt (GM) model \cite{iw, swr,dkp}, the
Gray-Scott (GS) model \cite{dek1,dek2,swr,chen}, the Schnakenberg
model \cite{rad}, a three-component RD system modeling gas-discharge
\cite{van}, the Brusselator model \cite{tzou_2}, a model for hot-spots
of urban crime \cite{tse}, and a general class of RD models
\cite{Nec5}. In these studies, a wealth of different analytical
techniques have been used, including the method of matched asymptotic
expansions, Lyapanov-Schmidt reductions, geometric singular
perturbation theory, and the rigorous renormalization approach of
\cite{dkp}. In contrast, for the case of a 2-D domain, there are only
a few studies of the dynamics of localized spot patterns by formal
asymptotic analysis (see \emph{e.g.}~\cite{kolok_ward,
  Kolokolnikov2009, Chen2011}), as the analytical techniques available
in 2-D are, to a large extent, very different in nature to those for
the simpler 1-D case.

There have been many numerical studies of RD patterns on the sphere
and other compact manifolds
(\emph{c.f.}~\cite{chaplain_spatio-temporal_2001, LV, barr,
  gjorgjieva_turing_2007, MMR:pnas2013,
  nagata_reaction-diffusion_2003, varea_turing_1999}), many of which
are motivated by specific problems in biological pattern formation for
both stationary and time-evolving surfaces
(\emph{c.f.}~\cite{painter_models_2000, kondo_reaction_1995,
  plaza2004growth}). Most prior analytical studies of pattern
formation on surfaces have been restricted to the sphere, and focus on
analyzing the development of small amplitude spatial patterns that
bifurcate from a spatially uniform steady-state at some critical
parameter value. Near this bifurcation point, weakly nonlinear theory
based on equivariant bifurcation theory and detailed group-theoretic
properties of the spherical harmonics have been used to derive and
analyze normal form amplitude equations characterizing the emergence
of these small amplitude patterns
(\emph{c.f.}~\cite{callahan_turing_2004,chossat,
  Matthews_1,Matthews_2, Pena, varea_turing_1999}). However, due to
the typical high degree of degeneracy of the eigenspace associated
with spherical harmonics of large mode number, these normal form
amplitude equations typically consist of a large coupled set of
nonlinear ODEs. These ODEs have an intricate subcritical bifurcation
structure, with weakly nonlinear patterns typically only becoming
stable past a saddle-node bifurcation point. As a result, the
preferred spatial pattern that emerges from an interaction of these
modes is difficult to predict theoretically. Moreover, although
equivariant bifurcation theory is able to readily predict the general
form of the coupled set of amplitude equations, the problem of
calculating the coefficients in these amplitude equations for specific
RD systems is rather intricate in general (see
\cite{callahan_turing_2004} for the case of the Brusselator).

In this paper and its companion \cite{rozada2014}, we propose an
alternative theoretical framework for analyzing RD patterns on the
sphere. In contrast to a weakly nonlinear framework, our theoretical
analysis is not based on an asymptotic closeness of parameters to a
Turing bifurcation point. Instead, it relies on an assumed large
diffusivity ratio between the two components in the system. In this
singularly perturbed limit, the Brusselator allows for the existence
of localized quasi-equilibrium spot-type patterns for a wide range of
parameters.

Related work on characterizing slow spot dynamics in
a 2-D planar domain was done previously for the Schnakenberg model
\cite{Kolokolnikov2009} and the Gray-Scott model \cite{Chen2011}. Our
analysis of slow spot dynamics on the sphere is rather more
complicated than that for the planar case since we must carefully
examine certain correction terms generated by the curvature of the
sphere.

We also note that an important motivation for this paper is to better
understand the connections that exists between the study of spot
patterns on the sphere in RD systems, with the apparently similar
study of point vortex motion on the sphere in Eulerian fluid
mechanics. For the latter problem, the positions of the point vortices
are similarly governed by a reduced dynamical system. This system
has been under intense study over the past three decades (see
\cite{Bogomolov,Newton1998,Newton2001,Newton2011,Boatto} and the references
therein).


\section{Two principal results for slow spot dynamics}\label{sec:main}

In this section, we present our main results for the slow dynamics of a
collection of localized spots for (\ref{full_all}) on the surface of the
unit sphere.  The first result, as originally derived in
\cite{rozada2014}, is an asymptotic result characterizing
quasi-equilibrium solutions of (\ref{full_all}) when $\eps\ll 1$. The
result is as follows:

\begin{result}[Quasi-Equilibria, (2.15) in \cite{rozada2014}] \label{thm:quasi} 
For $\eps \to 0$, the leading order uniformly valid quasi-equilibrium
solution to \eqref{full_all} is described by an outer solution, valid
away from the spots, and inner core solutions near each of the $N$
spots centered at $\bm{x}=\bm{x}_j$ for $j = 1, \ldots, N$. These
solutions are
\begin{equation}\label{quasi_eq}
  u_\unif \sim \eps^2 \E + \sum_{i=1}^N U_{i,0}\left(\frac{|\bm{x} -
    \bm{x}_i|}{\eps}\right)\,, \qquad v_\unif \sim \sum_{i=1}^N S_i
  L_i(\bm{x}) - 4\pi R\E + \overline{v}\,,
\end{equation}
where $L_i(\bm{x}) \equiv \log|\bm{x} - \bm{x}_i|$, $R \equiv
\frac{1}{4\pi}(\log 4 - 1)$, and $\overline{v}$ is a constant. The leading-order radially symmetric inner core solution, $U_{i,0}$, is defined on
the tangent plane to the sphere near the spot at $\bm{x} = \bm{x}_i$,
and is found by numerical computation of the BVP
\eqref{U0V0eqn}. In (\ref{quasi_eq}), the spot strengths, $S_i$ for
$i=1,\ldots,N$, satisfy the nonlinear algebraic system
\begin{equation}
\N(\bm{S}) \equiv \Bigl[ \I - \nu (\I - {\mathcal E}_0)\mathcal{G}\Bigr] 
\bm{S} + \nu (\I - {\mathcal E}_0) \bm{\chi}(\bm{S}) - \frac{2\E}{N} 
\bm{e}=\bm{0} \,.  \label{constraint}
\end{equation}
Here $\I$ is $N\times N$ identity matrix, $({\mathcal E}_0)_{ij} =
\frac{1}{N}$, $(\bm{S})_i = S_i$, $(\bm{\chi}(\bm{S}))_i = \chi(S_i)$,
$(\mathcal{G})_{ij} = L_i(\bm{x}_j)$ for $i\neq j$ and
$(\mathcal{G})_{ii} = 0$, $(\bm{e})_i = 1$, and
$\nu={-1/\log\eps}$. The values of $\chi(S_i)$ are found by
numerically solving the leading-order inner system \eqref{U0V0eqn}
(see Fig.~\ref{UVchi}). In terms of the spot strengths, the constant
$\overline{v}$ in (\ref{quasi_eq}) is
\begin{equation}\label{vc}
  \overline{v}=\frac{2\E}{\nu N} + 4\pi R \E + \frac{1}{N}
  \left[  \bm{e}^T \bm{\chi} - \bm{e}^T \mathcal{G} \bm{S} \right] \,.
\end{equation}
\end{result}

For a fixed configuration of spot locations, the linear stability of
such quasi-equilibrium solutions to ${\mathcal O}(1)$ time-scale
instabilities was investigated in \cite{rozada2014}. There, it was
found that, depending on the range of $\E$, $\tau$, and $f$, such
instabilities can lead to either spot self-replication events, a
spot-annihilation phenomena, or temporal oscillations of a spot
profile. These instabilities are discussed in detail in \S
\ref{sec:qe_instab}. For $\E={\mathcal O}(1)$, our analysis in \S
\ref{sec:qe_instab} shows that, to leading order in $\nu$,
spot-patterns for which $S_j={\mathcal O}(1)$, for all $j=1,\ldots,N$,
are linearly stable on an ${\mathcal O}(1)$ time-scale provided that
$S_j<\Sigma_2(f)$ for all $j=1,\ldots,N$, where $\Sigma_2$ is referred
to as the spot self-replication threshold.

However, in those parameter range where these $\Oh(1)$ time-scale
instabilities are absent, the main result of this paper is to show
that the quasi-equilibrium solution of (\ref{quasi_eq}) characterizes
the slow dynamics of a localized spot pattern for (\ref{full_all}) on
the longer time scales of $\Oh(\eps^{-2})$. On this long time-scale,
the slow dynamics of the centers of a collection of $N$ spots is
characterized as follows:

\begin{result}[Slow spot dynamics] \label{thm:dyn} Let $\eps\to 0$. 
Provided that there are no $\Oh(1)$ time-scale instabilities of the
quasi-equilibrium spot pattern, the time-dependent spot locations,
$\bm{x}_j = (\cos\phi_j\sin\theta_j, \sin\phi_j\sin\theta_j,
\cos\theta_j)^T$, vary on the slow time-scale $\sigma = \eps^2 t$, and
satisfy the differential algebraic system (DAE):
\begin{subequations}\label{slow_dyn}
\begin{equation}
\dd{\theta_j}{\sigma} = -\frac{2}{\mathcal{A}_j} \alpha_1(\bm{x}_j) \,,
\qquad \sin\theta_j \dd{\phi_j}{\sigma} =
-\frac{2}{\mathcal{A}_j} \alpha_2(\bm{x}_j) \,, \qquad j=1,\ldots,N \,,
 \label{slow_dyn_a}
\end{equation}
where ${\mathcal A}_j={\mathcal A}(S_j;f)$ is
a nonlinear function of $S_j$ defined via an integral in (\ref{asolv})
(see Fig.~\ref{fig:core2}), and
\begin{equation}
\begin{pmatrix}
\alpha_1(\bm{x}_j) \\ \alpha_2(\bm{x}_j)
\end{pmatrix}
 \equiv \sum_{\substack{i=1 \\ i \neq j}}^N S_i
\begin{pmatrix}
\pd{L_i(\bm{x})}{\theta} \\
\frac{1}{\sin\theta_j} \pd{L_i(\bm{x})}{\phi} \end{pmatrix}\Biggr
\rvert_{\phi=\phi_j,\theta=\theta_j}  \,.
\end{equation}
The spot strengths $S_j$, for $j=1,\ldots,N$, are coupled to the slow
dynamics by (\ref{constraint}).
\end{subequations}
\end{result}

It is convenient to express the slow dynamics of the spot locations in
a more explicit form. To do so, we use the cosine law
$|\bm{x}-\bm{x}_i|^2=2(1-\cos\gamma_i)$ to write $L_i$ in terms of
spherical coordinates as
\begin{equation*}
  L_i=\frac{1}{2}\log\left[1-\cos\gamma_i \right] + \frac{1}{2}\log{2}
   \,, \qquad
  \cos\gamma_i = \cos\theta \cos\theta_i + \sin\theta \sin\theta_i
  \cos(\phi-\phi_i) \,,
\end{equation*}
where $\gamma_i=\gamma_i(\phi,\theta)$ is the angle between $\bm{x}$ and
$\bm{x}_i$. By using this form for $L_i$, (\ref{slow_dyn}) becomes
\begin{subequations} \label{slow_dyn_e}
\begin{align}
\dd{\theta_j}{\sigma} &= -\frac{1}{\mathcal{A}_j} 
\sum_{\substack{i=1 \\ i \neq j}}^N \left(\frac{S_i}{1-\cos\gamma_{ij}}\right)
  \Bigl[\sin\theta_j\cos\theta_i-\cos\theta_j\sin\theta_i\cos(\phi_j-\phi_i)
 \Bigr] \,, \\
\qquad \sin\theta_j \dd{\phi_j}{\sigma} &=
-\frac{1}{\mathcal{A}_j} \sum_{\substack{i=1 \\ i \neq j}}^N 
\left(\frac{S_i}{1-\cos\gamma_{ij}}\right)\Bigl[ \sin\theta_i \sin(\phi_j-\phi_i)\Bigr] \,,
\end{align}
\end{subequations}
for $j=1,\ldots,N$, where $\gamma_{ij}\equiv\gamma_i(\phi_j,\theta_j)$ is the
angle between $\bm{x}_i$ and $\bm{x}_j$.

As an alternative to (\ref{slow_dyn_e}), we can also write
(\ref{slow_dyn}) in terms of cartesian coordinates. Writing ${\bm
  x}_j$ as a column vector, and letting $T$ denote transpose, we will
show in \S\ref{sec:asym} that (\ref{slow_dyn_e}) is equivalent to
\begin{equation}\label{slow_cart}
  \dd{\bm{x}_j}{\sigma} = \frac{2}{\mathcal{A}_j} \left( \I - {\mathcal Q}_j
\right) \sum_{\substack{i=1 \\ i \neq j}}^N  \frac{S_i \bm{x}_i}{|\bm{x}_i-\bm{x}_j|^2}
 \,, \qquad {\mathcal Q}_j\equiv \bm{x}_j \bm{x}_j^T \,, 
\qquad j=1,\ldots,N \,.
\end{equation}

\section{Asymptotic derivation of the slow spot dynamics}\label{sec:asym}

Our aim in this section is to construct a localized quasi-equilibrium
spot pattern solution for the system \eqref{full_all} in the limit
$\eps \to 0$. Such solutions consist of two parts. The first consists of an outer region, where
the solution varies slowly according to 
\begin{equation}
u_\out \sim \eps^2 \E \quad \text{and} \quad \dell_S v_\out \sim -\eps^{-2} \Fv(u_\out,v_\out)\sim -\E.
\end{equation}
The second part of the solution consists of localized inner regions of spatial extent ${\mathcal O}(\eps)$ near each of the spots centered at $\bm{x}=\bm{x}_j$, where $\bm{x}_j =
(\cos\phi_j \sin \theta_j, \sin\phi_j \sin\theta_j, \cos\theta_j)^T$,
for $j = 1, \ldots N$.

\subsection{Plan of action}

The asymptotic analysis presented below is necessarily detailed and technical, so let us first outline the three main steps of the procedure.

(Step 1) We first apply a dominant balance argument and argue that the
centers of the spots will move slowly on a time scale $\sigma$ defined
by $\sigma = \eps^2 t$, so that $\bm{x}_j=\bm{x}_j(\sigma)$. In the
inner region near the $j^\textrm{th}$ spot we introduce the local
coordinates $\bm{s} =(s_1, s_2)^T$ defined by
\begin{equation} \label{s1s2}
s_1 \equiv \eps^{-1} \left[\theta - \theta_j(\sigma)\right] \,, \qquad
s_2 \equiv \eps^{-1} \sin\theta_j \left[\phi-\phi_j(\sigma)\right] \,, \qquad
  \sigma=\eps^2 t \,.
\end{equation}
By re-scaling into the inner region, we develop the zeroth (leading) and first-order 
equations for the two inner solutions,
$\uin=U_j(\bm{s},\sigma)$ and $\vin=V_j(\bm{s},\sigma)$ near the
$j^\text{th}$ spot. The leading-order inner problem is precisely the
same as in \cite{rozada2014}. The first correction, however, is new,
and is necessary in order to establish the dynamics.

(Step 2) We return to the outer region and develop a uniformly-valid
outer solution which includes the logarithmic behaviour near the inner
region (expressed as a sum of Green's functions) and the unknown
source strengths, $S_j$. Matching the inner and outer solutions at
leading-order gives Principal Result \ref{thm:quasi}, \emph{i.e.} a
nonlinear algebraic equation for $S_j$ for a known set of
$\bm{x}_j$. Both Steps 1 and 2 are nearly the same as in
\cite{rozada2014}. The only difference is that the matching procedure
of Step 2 requires derivation of higher-order terms that are used
later.

(Step 3) The derivation of the governing equation for the spot
locations now follows from matching the inner and outer solutions at
first order, and applying a solvability condition. A key
difficulty that confronts us in this step is that the higher-order
matching between inner and outer solutions requires not only matching
the inter-spot interactions and slow dynamics of the patterns, but
also the corrections that arise due to the projection of the spherical
geometry onto the local tangent plane approximation.

Before proceeding with these three steps, we first establish the
following lemma that explains how the outer coordinate, $\bm{x}$, can
be re-written in terms of the inner coordinate, $\bm{s}$. The proof is
presented in \ref{proof:coord}.
\begin{restatable}[Tangent plane transformation]{lemma}{lemmaplane}\label{lemma:coord} 
Suppose that $\theta_j\in (0,\pi)$. Then, for 
$|{\bm x}-{\bm x}_j|={\mathcal O}(\eps)$ and $|\bm{s}|={\mathcal
    O}(1)$, we have
\begin{subequations}\label{map}
\begin{equation}
 \bm{x}-{\bm x}_j = \eps \bm{J}_j {\bm s} + {\mathcal O}(\eps^2)
\,, \qquad  |{\bm x}-{\bm x}_j| \sim \eps \rho +  \frac{\eps^2}{2\rho}
 s_1 s_2^2 \cot\theta_j \,, \qquad \rho\equiv \left(s_1^2+s_2^2\right)^{1/2}
 \,, \label{map_1}
\end{equation}
where $\bm{s}\equiv (s_1,s_2)^T$ and $\bm{J}_j$ is the $3\times 2$ matrix
defined by
\begin{equation}
   \bm{J}_j^T \equiv \begin{pmatrix}
       \cos\phi_j \cos\theta_j & \sin\phi_j\cos\theta_j & -\sin\theta_j \\
      -\sin\phi_j   & \cos\phi_j  &  0 \\
                     \end{pmatrix}\,. \label{map_3}
\end{equation}
\end{subequations}
\end{restatable}

\subsection{(Step 1) Governing equations near the spots}

We begin by re-scaling the governing equations near the $j^\text{th}$
spot and proceed to develop the first two orders. First, we write
$\uin=U_j(\bm{s},\sigma)$ and $\vin=V_j(\bm{s},\sigma)$, and expand
\begin{equation}
 U_j(\bm{s}, \sigma) = \sum_{n=0}^\infty \eps^n U_{jn} \,,\qquad
 V_j(\bm{s}, \sigma) = \sum_{n=0}^\infty \eps^n V_{jn}\,.\label{uv_inn}
\end{equation}
In addition, upon introducing (\ref{s1s2}) into (\ref{laplace}) and the
time derivative, we obtain for $\eps\ll 1$ that
\begin{subequations}\label{l_new}
\begin{equation}
\dell_S = \frac{1}{\eps^2} \dell_{(s_1,s_2)} + \frac{1}{\eps} \enn_1 + 
{\mathcal O}(1)\,, \qquad \pd{}{t} = \eps \tee_1 + \eps^2 
\pd{}{\sigma} \,,
\end{equation}
where we have defined the additional operators,
\begin{gather}
\nabla_{(s_1,s_2)} \equiv \left(\pd{}{s_1},\pd{}{s_2}\right)\,, \qquad
\dell_{(s_1,s_2)} \equiv \pd{^2}{s_1^2} + \pd{^2}{s_2^2}\,, \\ \enn_1
\equiv \cot \theta_j \left( \pd{}{s_1} - 2s_1 \pd{^2}{s_2^2}\right) \,,
\qquad \tee_1 \equiv - \left(\dtheta_j, \dphi_j \sin\theta_j\right)
\cdot \nabla_{(s_1,s_2)} \,. \label{enntee}
\end{gather}
\end{subequations}
Here the overdot indicates derivatives with respect to $\sigma$, We
substitute (\ref{uv_inn}) and (\ref{l_new}) into (\ref{full_all}), and
equate powers of $\eps$ to obtain inner problems near
$\bm{x}=\bm{x}_j$. To leading order, on $\bm{s}\in\R^2$ we obtain the
same set of equations as presented in \cite{rozada2014} [\emph{c.f.}
  their (2.1)]:
\begin{subequations} \label{U0V0}
\begin{align}
\dell_{(s_1,s_2)} U_{j0} - U_{j0} + f U_{j0}^2 V_{j0} &= 0\,, \\
\dell_{(s_1,s_2)} V_{j0} + U_{j0} - U_{j0}^2 V_{j0} &= 0 \,.
\end{align}
\end{subequations}

At next order, and labelling $\bm{U}_{j1}\equiv (U_{j1},V_{j1})^T$ and
$\bm{U}_{j0}\equiv (U_{j0},V_{j0})^T$, we find on $\bm{s}\in \R^2$
that
\begin{equation}\label{U1V1}
\Lop \bm{U}_{j1} \equiv \dell_{(s_1,s_2)} \bm{U}_{j1} + {\mathcal M}_j 
\bm{U}_{j1} =  
 - \enn_1 \bm{U}_{j0} + \begin{pmatrix}
                         \tee_1 U_{j0} \\
                         0  \\
                     \end{pmatrix} \,,  \qquad
{\mathcal M}_j\equiv  
\begin{pmatrix}
-1 + 2f U_{j0} V_{j0}  &  f U_{j0}^2 \\
1- 2U_{j0} V_{j0}      &  - U_{j0}^2 \\
\end{pmatrix} \,.
\end{equation}
Indeed, it is the above set of equations that will be used to
establish the dynamics of the spots.

\subsection{(Step 2) The leading-order inner problem and initial matching}

We now move on to solve the leading-order inner problem \eqref{U0V0}
and perform the leading-order matching between inner and outer
solutions. This procedure will lead to Principal
Result~\ref{thm:quasi}.

First, we seek a radially symmetric solution to (\ref{U0V0}) with
matching conditions,
\begin{equation}
U_{j0} \to 0 \quad \text{and} \quad V_{j0}\sim S_j\log\rho \quad 
\text{as $\rho\to \infty$,}  
\end{equation}
where $\rho\equiv (s_1^2+s_2^2)^{1/2}$ is the distance from the spot
along the tangent plane, and $S_j$, referred to as the spot strength,
is a parameter to be determined (cf.~\cite{rozada2014}). Note that since
$u_\out ={\mathcal O}(\eps^2)$ in the outer region, the
far-field behavior of $U_{j0}$ matches with the outer solution.

As such, in (\ref{U0V0}) we set $U_{j0} = U_{j0}(\rho)$ and $V_{j0}
=V_{j0}(\rho)$. In terms of $\dell_\rho \equiv \partial_{\rho\rho} +
\rho^{-1} \partial_\rho$, \eqref{U0V0} reduces to the following BVP
system on $0<\rho<\infty$:
\begin{subequations} \label{U0V0eqn}
\begin{gather}
\dell_\rho U_{j0} - U_{j0} + f U_{j0}^2 V_{j0} = 0 \,, \qquad
\dell_\rho V_{j0}  + U_{j0} - U_{j0}^2 V_{j0} = 0 \,, \label{U0V0eq} \\
U_{j0}^{\prime}(0) = V_{j0}^{\prime}(0)=0 \,, \quad
U_{j0} \to 0 \,, \quad V_{j0} \sim S_j \log \rho + \chi + o(1) \,
\text{ as $\rho \to \infty$}. \label{V0bc}
\end{gather}
\end{subequations}
In general, the solution of the above problem must be computed
numerically, and we have included in \ref{appendix:core} additional
details and figures for such computations. In particular,
  the far-field constant $\chi=\chi(S_j;f)$, which is needed for the
  slow dynamics, must be computed numerically. Upon integrating
\eqref{U0V0eq} for $V_{j0}$, we obtain the identity $S_j =
\int_0^\infty (U_{j0}^2 V_{j0} - U_{j0}) \rho \, \de{\rho}$.

Next, we relate the outer solution for $v$, valid away from the spots,
to the inner solution $V_{j0}$. We first use the leading-order
uniformly valid solution for $u$, given by $u_\unif \sim \eps^2 \E +
\sum_{i=1}^N U_{i0}$, to calculate $H(u,v)$, defined in
(\ref{full_all}), in the sense of distributions as
\begin{equation}
\eps^{-2} (u - u^2 v) \sim \E + 2\pi \sum_{i=1}^N 
\left[\int_0^\infty (U_{i0} - U_{i0}^2 V_{i0})\rho \, \de{\rho}\right] 
\sim \E - 2\pi \sum_{i=1}^N S_i \delta(\bm{x} - \bm{x}_i)\,. 
\label{delt1}\end{equation}
In this way, we obtain from (\ref{full_all}) that the leading-order
outer approximation for $v$ satisfies
\begin{equation} \label{vunifeq}
\dell_S v = - \E + 2\pi \sum_{i=1}^N S_i \delta(\bm{x} - \bm{x}_i) \,, \quad
 \text{where} \quad \sum_{i = 1}^N S_i = 2\E \,.
\end{equation}

The solution to \eqref{vunifeq}, subject to smoothness conditions at
the two poles, can be written in terms of the unique
source-neutral Green's function $G(\bm{x};\bm{x}_i)$ defined by
\begin{equation}\label{Gfeq}
\dell_S G = \frac{1}{4\pi} - \delta(\bm{x} - \bm{x}_i)\, \quad \text{and} \quad
\int_\text{unit sphere} G \, \de{x} = 0\,.
\end{equation}
The well-known solution to \eqref{Gfeq} is
\begin{equation} \label{Gexact}
G(\bm{x}; \bm{x}_i) = -\frac{1}{2\pi} L_i(\bm{x}) + R\,, \qquad R = 
\frac{1}{4\pi}[\log 4 - 1]\, \qquad L_i(\bm{x}) \equiv \log|\bm{x} - \bm{x}_i|
\,.
\end{equation}
Thus, in terms of $G$, the solution to (\ref{vunifeq}) is given by 
\begin{equation} \label{vunifguess}
v =  -2\pi \sum_{i=1}^N S_i G(\bm{x}; \bm{x}_i) + \overline{v} = 
\sum_{i=1}^N S_i L_i(\bm{x}) - 4\pi R \E + \overline{v}\,,
\end{equation}
for some constant $\overline{v}$ to be determined below from matching to each
inner solution $V_{j0}$.

To determine the spot strengths, $S_j$ for $j = 1, \ldots, N$, and 
the unknown constant $\overline{v}$, we  match the outer and inner solutions
for $v$. We expand the outer solution in (\ref{vunifguess}) as 
$\bm{x}\to\bm{x}_j$ to obtain
\begin{equation*}
   v\sim S_j \log|\bm{x}-\bm{x}_j| - 4\pi R\E + \overline{v} +
\sum_{\substack{i=1 \\ i \neq j}}^N S_i L_{ij}  + 
\sum_{\substack{i=1 \\ i \neq j}}^N S_i \nabla_{\bm{x}} L_i \bigr\rvert_{\bm{x}=\bm{x}_j}
  \cdot (\bm{x}-\bm{x}_j) + \cdots \,,
\end{equation*}
where $L_{ij}\equiv \log|\bm{x}_i-\bm{x}_i|$.  Then, we use
(\ref{map_1}) to write this expression in the inner variable $\bm{s}$
as
\begin{equation}\label{match_1}
   v\sim S_j \left[\log\eps + \log\rho + \frac{\eps}{2\rho^2}
     s_1 s_2^2 \cot\theta_j \right] - 4\pi R\E + \overline{v} +
   \sum_{\substack{i=1 \\ i \neq j}}^N S_i L_{ij} + \eps
   \sum_{\substack{i=1 \\ i \neq j}}^N S_i J_{j}^T \nabla_{\bm{x}} L_i
  \bigr\rvert_{\bm{x}=\bm{x}_j} \cdot \bm{s} \,,
\end{equation}
where $\rho=(s_1^2+s_2^2)^{1/2}$ and $\bm{J}_j$ is defined in
(\ref{map_3}). In contrast, the far-field behavior of the $j^\textrm{th}$ inner
solution is $V_{j}\sim S_{j}\log\rho + \chi(S_j) + \eps V_{j1}+\cdots$. To
match the far-field behavior of this inner solution with (\ref{match_1}),
we require that
\begin{subequations}\label{match_2}
\begin{gather}
   S_j \log\eps - 4\pi R \E + \overline{v} + 
   \sum_{\substack{i=1 \\ i \neq j}}^N S_i L_{ij} =\chi(S_j) \,, \qquad
   j=1,\ldots,N \,, \label{match_2a}
  \\
  V_{j1} \sim \frac{S_j}{2\rho^2}
     s_1 s_2^2 \cot\theta_j  + \sum_{\substack{i=1 \\ i \neq j}}^N S_i J_{j}^T 
  \nabla_{\bm{x}} L_i \bigr\rvert_{\bm{x}=\bm{x}_j} \cdot \bm{s}\,, 
\,\,\, \text{ as $|\bm{s}|\to \infty$}\,; \qquad j=1,\ldots,N \,. 
\label{match_2b}
\end{gather}
\end{subequations}

From (\ref{match_2a}), and noting the constraint in (\ref{vunifeq}),
we obtain that $S_j$ for $j=1,\ldots,N$ and $\overline{v}$ satisfy the
$N+1$ dimensional nonlinear algebraic system
\begin{subequations}\label{match_3}
\begin{equation}
   S_j + \nu \chi(S_j) - \nu \sum_{\substack{i=1 \\ i \neq j}}^N S_i L_{ij} =
  \overline{v}_c \,, \quad j=1\,,\ldots,N \,; \qquad 
\sum_{i = 1}^N S_i = 2\E \,, \label{match_3a}
\end{equation}
where $\nu$, $L_{ij}$, and $\overline{v}_c$ are defined by
\begin{equation}
   \nu \equiv {-1/\log\eps} \,, \qquad L_{ij}=\log|\bm{x}_i-\bm{x}_j|\,, 
  \qquad \overline{v}\equiv\frac{\overline{v}_c}{\nu}+ 4\pi R \E \,.
  \label{match_3b}
\end{equation}
\end{subequations}
By writing (\ref{match_3a}) in matrix form, we then eliminate the
constant $\overline{v}_c$ to derive that the spot strengths
satisfy the nonlinear algebraic system in (\ref{constraint}). In terms
of the spot strengths, the constant $\overline{v}$ is given in
(\ref{vc}). This completes the derivation of Principal Result
\ref{thm:quasi}.

Now before proceeding to the final step and deriving of the dynamical
equations for the spots, let us draw the reader's attention to the
far-field behaviour of $V_{j1}$ in \eqref{match_2b}. The work of this
section that had led to Principal Result 1 is identical to that shown
in \cite{rozada2014}, with the exception of the details surrounding
this higher-order far-field behaviour.

\subsection{(Step 3) The solvability condition and higher-order matching}

To derive the result in Principal Result \ref{thm:dyn} for the slow
spot dynamics, we must analyze the first-order inner problem (\ref{U1V1})
subject to the far-field condition (see (\ref{match_2b})) that
\begin{equation}\label{U1V1_inf}
  \bm{U}_{j1} \equiv \begin{pmatrix}
                U_{j1} \\ 
                V_{j1} \\
             \end{pmatrix} \sim 
                     \begin{pmatrix}
                     0 \\
                   \frac{S_j}{2\rho^2}
     s_1 s_2^2 \cot\theta_j  + \sum_{\substack{i=1 \\ i \neq j}}^N S_i J_{j}^T 
  \nabla_{\bm{x}} L_i \bigr\rvert_{\bm{x}=\bm{x}_j} \cdot \bm{s} \\
              \end{pmatrix} \,, \qquad  \mbox{as} \quad \rho=
|\bm{s}|\to \infty \,.
\end{equation}
Of the four inhomogeneous terms in (\ref{U1V1}) and (\ref{U1V1_inf}),
the forcing term $\enn_1 \bm{U}_{j0} $ in (\ref{U1V1}) and the term
$S_j s_1s_2^2 \cot \theta_j/(2\rho^2)$ in (\ref{U1V1_inf}) correspond
to corrections to the leading-order tangent plane approximation to the
sphere at $\bm{x}=\bm{x}_j$.  These correction terms are present even for
the case of a single stationary spot solution. In contrast, the
two remaining inhomogeneous terms in (\ref{U1V1}) and (\ref{U1V1_inf})
result either from inter-spot interactions or from the time operator,
$\tee_1$, applied to $U_{j0}$.

With this motivation, we seek a decomposition for $\bm{U}_{j1}$ into a
``static'' component, reflecting correction terms to the tangent plane
approximation, and a ``dynamic'' component resulting from inter-spot
interactions. This decomposition of the solution $\bm{U}_{j1}$ to
(\ref{U1V1}) with (\ref{U1V1_inf}) has the form
\begin{equation}
  \bm{U}_{j1} \equiv \UVb = \Ueq + \Udyn \,, \qquad 
 \Ueq \equiv \UVeq \,, \qquad   \Udyn \equiv \UVdyn \,, \label{u1_decomp}
\end{equation}
where, in terms of the operator $\Lop$ of (\ref{U1V1}), $\Ueq$ satisfies
\begin{equation} \label{Ueq_eqn}
\Lop\Ueq = -\enn_1 \bm{U}_{j0} \,, \quad \bm{s}\in \R^2\,; \qquad
\Ueq \sim   \begin{pmatrix}
0 \\ 
\frac{S_j}{2\rho^2} s_1 s_2^2 \cot \theta_j
\end{pmatrix} \,,\quad \mbox{as} \quad |\bm{s}| \to \infty \,.
\end{equation}
In contrast, the dynamic component $\Udyn$ is taken to satisfy
\begin{subequations}
\begin{equation} \label{Udyn_eqn}
\Lop\Udyn = \begin{pmatrix} \tee_1 U_{j0} \\ 0\end{pmatrix} \,,
 \quad \bm{s}\in \R^2\,; \qquad
  \Udyn \sim 
\begin{pmatrix}
0 \\ 
\bm{\alpha} \cdot \bm{s} \end{pmatrix} \,, 
\quad \mbox{as} \quad |\bm{s}| \to \infty \,.
\end{equation}
Here $\bm{\alpha}$, identified from the second term in
(\ref{U1V1_inf}), is given by
\begin{equation}\label{alp_def}
 \bm{\alpha}\equiv \sum_{\substack{i=1 \\ i \neq j}}^N S_i J_{j}^T 
  \nabla_{\bm{x}} L_i \bigr\rvert_{\bm{x}=\bm{x}_j} = 
  \sum_{\substack{i=1 \\ i \neq j}}^N  S_i 
  \begin{pmatrix}
  \pd{L_{i}}{\theta} \\
  \frac{1}{\sin\theta_j}\pd{L_{i}}{\phi}
  \end{pmatrix}\Biggr\rvert_{\phi=\phi_j,\theta=\theta_j}\,.
\end{equation}
\end{subequations}
Next, we show that a particular solution to (\ref{Ueq_eqn}) can be
identified analytically. The proof is presented in \ref{proof:tangent}.

\begin{lemma}[Static component of first-order inner solution] \label{lemma:tangent} Suppose that $U_0(\rho)$ and $V_0(\rho)$, with
$\rho=(s_1^2+s_2^2)^{1/2}$, are radially symmetric solutions to
\begin{subequations} \label{part_0}
\begin{gather}
\dell_{(s_1,s_2)} U + F(U,V) = 0 \,, \qquad
\dell_{(s_1,s_2)} V + H(U,V) = 0 \,, \quad 0<\rho<\infty \,, \\
U \to 0 \,, \qquad V \sim S_j \log \rho + \chi + o(1) \,,
\quad \mbox{as} \quad \rho \to \infty \,,
\end{gather}
\end{subequations}
where $\dell_{(s_1,s_2)}\equiv \partial_{s_1s_1}+\partial_{s_2s_2}$. Then,
consider the linearized problem for $\bm{U}_1$ on $\bm{s}\in \R^2$
formulated as
\begin{subequations} \label{part_1}
\begin{gather}
\Lop \bm{U}_{1} \equiv \dell_{(s_1,s_2)} \bm{U}_{1} + {\mathcal M} 
\bm{U}_{1} =  - \cot\theta_j \left( \bm{U}_{0s_1} - 2s_1 \bm{U}_{0s_2s_2}
 \right)\,, \label{part_1a} \\
 {\mathcal M} \equiv \begin{pmatrix}
                  F_{U}   &   F_{V} \\
                  H_{U}   &   H_{V} \\
                      \end{pmatrix}\Biggr\rvert_{(U,V)=(U_0,V_0)}
\hspace*{-4.5em}, \hspace*{4.5em}\qquad  
\bm{U}_1 \sim \begin{pmatrix}
            0 \\ 
            \frac{S_j}{2\rho^2} s_1 s_2^2 \cot \theta_j
          \end{pmatrix} \text{ as $|\bm{s}| \to \infty$}\,.
\label{part_1b}
\end{gather}
\end{subequations}
Here $\bm{U}_1\equiv (U_1,V_1)^T$ and $\bm{U}_0\equiv (U_0,V_0)^T$,
Then, a solution to (\ref{part_1}) is
\begin{equation}
   \bm{U}_1 = -\frac{s_2^2}{2} \cot\theta_j  
\left(\partial_{s_1} \bm{U}_{0} \right) +
   \cot\theta_j s_1 s_2 \left(\partial_{s_2} \bm{U}_0 \right) \,. \label{part:sol}
\end{equation}
\end{lemma}

By applying this lemma to (\ref{Ueq_eqn}) we identify the static component
as
\begin{equation}
  \Ueq = -\frac{s_2^2}{2} \cot\theta_j \partial_{s_1}\bm{U}_{j0} +
   \cot\theta_j s_1 s_2 \partial_{s_2} \bm{U}_{j0} \,, \label{u1eq:sol}
\end{equation}
where $\bm{U}_{j0}=(U_{j0},V_{j0})^T$ satisfies (\ref{U0V0eqn}). The
key implication of this lemma is that the determination of $\Ueq$ is
independent of the particular form of the reaction kinetics. As such,
this lemma can be readily used for analyzing the dynamics of localized
spot patterns for other RD systems.

The final step in the analysis of the slow dynamics is to impose a
solvability condition on the dynamic component (\ref{Udyn_eqn}) for
$\Udyn$.  Since $\Lop \left(\partial_{s_i} \bm{U}_0\right)=0$ for
$i=1,2$, the dimension of the nullspace of the adjoint
$\Lop^{\star}$ is two-dimensional. For the homogeneous adjoint problem
\begin{equation}\label{adj}
  \Lop^{\star} \bm{\Psi} \equiv \dell_{(s_1,s_2)} \bm{\Psi} + {\mathcal M}_j^T 
\bm{\Psi} = 0\,,
\end{equation}
we look for separable solutions of the form
\begin{equation}\label{form:adj}
 \bm{\Psi}(\rho,\omega) = \bm{P}(\rho)T(\omega) \,, \qquad
  \bm{P} \equiv \begin{pmatrix}
 P_1(\rho) \\ P_2(\rho)
 \end{pmatrix} \,, \qquad \dell_\rho\equiv \partial_{\rho\rho} + 
\frac{1}{\rho}\partial_\rho \,,
\end{equation}
for local polar coordinates $\bm{s} = (\rho\cos\omega, \rho\sin\omega)^T$ where
$T(\omega)=\lbrace{\cos\omega,\sin\omega\rbrace}$. Thus, $\bm{P}$ 
satisfies
\begin{equation}\label{peq}
\dell_\rho \bm{P} - \frac{1}{\rho^2}\bm{P} + {\mathcal M}_j^T \bm{P} = 0 \,,
\end{equation}
with $\bm{P}\to \bm{0}$ as $\rho\to \infty$. To determine the
appropriate far-field behavior for $\bm{P}$, we observe that since
$U_{j0}\to 0$ exponentially as $\rho\to \infty$, then ${\mathcal M}_j$
from (\ref{U1V1}) satisfies
\begin{equation*}
   {\mathcal M}_j^T\to \begin{pmatrix}
                           -1 & 1 \\
                            0 & 0 \\
                        \end{pmatrix}\,, \quad\text{ as $\rho \to \infty$}\,.
\end{equation*}
As such, the solution $P_2$ to (\ref{peq}) satisfies 
$P_2={\mathcal O}(\rho^{-1})$ as $\rho\to \infty$, consistent with the
decaying solution to $\dell_\rho P_2 - \rho^{-2} P_2=0$. We normalize
the eigenfunction by imposing that $P_2\sim {1/\rho}$ as $\rho\to \infty$.
With this normalization, and from the limiting form of the first row of 
${\mathcal M}_j^T$ for $\rho\gg 1$, we conclude from (\ref{peq}) that
$P_1\sim {1/\rho}$ as $\rho\to \infty$. In this way, we solve (\ref{peq})
subject to $\bm{P}\sim ({1/\rho},{1/\rho})^T$ as $\rho\to \infty$.

We now impose a solvability condition on the solution to
(\ref{Udyn_eqn}) with $\bm{\Psi}_1=\bm{P} T(\omega)$. We let $B_\sigma
\equiv \lbrace{ \bm{s}: \lvert \bm{s}\rvert\leq\sigma\rbrace}$.  By
applying Green's second identity to $\Udyn$ and $\bm{\Psi}_1$ we
obtain
\begin{equation}
\lim_{\sigma\to\infty} \int_{B_\sigma} \left[ \bm{\Psi}_1^T 
 \Lop \Udyn - (\Udyn)^T \Lop^{\star} \bm{\Psi}_1 \right] \, d{\bm s} = 
     \lim_{\sigma\to\infty} \int_{0}^{2\pi} \left( \bm{\Psi}_1^T \partial_\rho
    \Udyn - (\Udyn)^T \partial_{\rho} \bm{\Psi}_1^T \right) 
  \biggr\rvert_{\rho=\sigma}  \sigma  \, \de{\omega} \,. \label{solv_1}
\end{equation}
We now use the limiting far-field asymptotic behavior
\begin{equation*}
    \Udyn \sim \begin{pmatrix}
                   0 \\
                \alpha_1 \rho \cos\omega + \alpha_2 \rho\sin\omega\\
                \end{pmatrix} \,, \qquad
    \bm{\Psi}_1 \sim 
                \begin{pmatrix}
                   {1/\rho} \\
                  {1/\rho} \\

           \end{pmatrix} T(\omega)\,, \quad\text{ as $\rho \to \infty$}\,,
\end{equation*}
to calculate the right hand-side of (\ref{solv_1}), labeled by $\Lambda$, as
\begin{equation}\label{solv_right}
 \Lambda \equiv \int_{0}^{2\pi} \left[  2\alpha_1\cos\omega +
  2\alpha_2 \sin\omega \right] T(\omega) \, \de{\omega} =\begin{cases}
2\pi \alpha_1 & \text{ if $T(\omega) = \cos\omega$}\\
2\pi \alpha_2 & \text{ if $T(\omega) = \sin\omega$}
\end{cases}\,.
\end{equation}
Then, by substituting the right hand-side of (\ref{Udyn_eqn}) into the
left hand-side of (\ref{solv_1}), and using $\partial_{s_1} U_{j0} =
U_{j0}^{\prime}(\rho)\cos\omega$ and
$\partial_{s_2}U_{j0}=U_{j0}^{\prime}(\rho)\sin\omega$, we obtain that
\begin{equation}\label{solv_left}
 \Lambda =  -\lim_{\sigma\to\infty} \int_{0}^{\infty} \int_{0}^{2\pi} P_1(\rho) \left[
  \theta_{j}^{\prime} U_{j0}^{\prime}(\rho) \cos\omega + \sin\theta_j \phi_j^{\prime}
    U_{j0}^{\prime}(\rho) \sin\omega \right] \rho T(\omega) \, \de{\rho} \,
  \de{\sigma} \,.
\end{equation}
Upon using the two forms $T(\omega)=\cos\omega$ and
$T(\omega)=\sin\omega$, (\ref{solv_left}) with (\ref{solv_right}) for
$\Lambda$, reduces to (\ref{slow_dyn_a}), where we have defined
${\mathcal A}_j={\mathcal A}(S_j;f)$ by
\begin{equation}\label{asolv}
  \mathcal{A}_j \equiv \int_0^\infty U_{j0}^{\prime}(\rho) P_1(\rho) \rho \, 
\de{\rho} \,,
\end{equation}
which appears in the ODE part of our result
  (\ref{slow_dyn}) for slow spot dynamics.  Then, by substituting the
second expression for $\bm{\alpha}=(\alpha_1,\alpha_2)^T$, as given in
(\ref{alp_def}), into (\ref{slow_dyn_a}) we obtain the slow dynamics
(\ref{slow_dyn}) as written in Principal Result \ref{thm:dyn}.

\begin{figure}[htb]
\begin{center}
\includegraphics{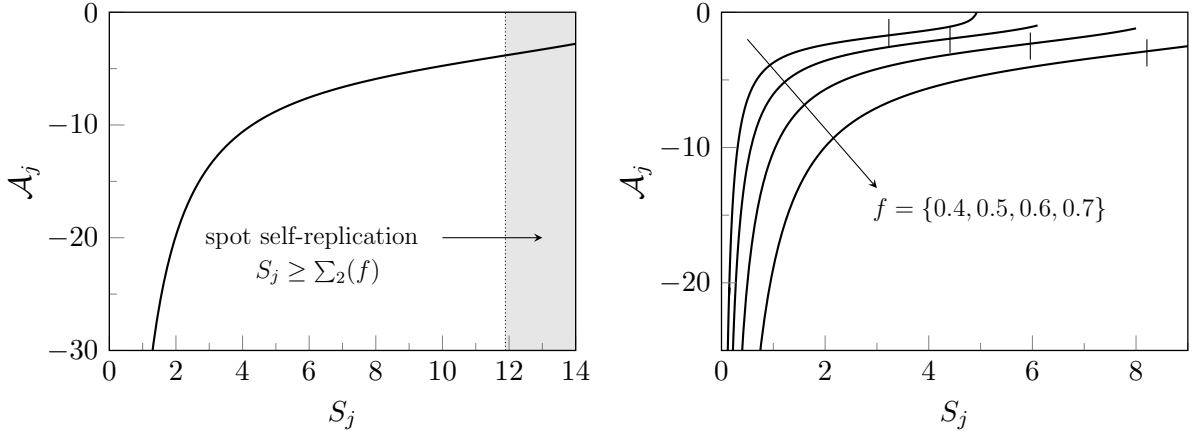}
\caption{Left: ${\mathcal A}_j$ versus $S_j$ for $f=0.3$. Right:
  ${\mathcal A}_j$ versus $S_j$ for $f=0.4$, $f=0.5$, $f=0.6$, and
  $f=0.7$, as shown. The thin vertical lines in these figures is the
  spot self-replication threshold $S_j=\Sigma_2(f)$ (see
  (\ref{self_rep})).  For $S_j>\Sigma_2(f)$, the quasi-equilibrium
  spot solution is linearly unstable on an ${\mathcal O}(1)$
  time-scale. On the range $0<S_j<\Sigma_2(f)$ we observe that
  ${\mathcal A}_j<0$. In this figure, the values of $f$ decrease in
  the direction of the arrow.}
\label{fig:core2}
\end{center}
\end{figure}

To implement (\ref{slow_dyn}), we must numerically compute ${\mathcal
  A}(S_j;f)$ from first solving the core problem (\ref{U0V0eqn}) for
$U_{j0}$ and then the adjoint problem (\ref{peq}) with far-field
behavior $\bm{P}\sim ({1\rho},{1/\rho})^T$ as $\rho\to \infty$. For
$f=0.3$, in the left panel of Fig.~\ref{fig:core2} we plot ${\mathcal
  A}_j$ versus $S_j$ for $f=0.3$. In the right panel of
Fig.~\ref{fig:core2} we plot ${\mathcal A}_j$ versus $S_j$ for
$f=0.4$, $f=0.5$, $f=0.6$, and $f=0.7$.

Finally, we show how (\ref{slow_cart}) follows from
(\ref{slow_dyn}). We first differentiate $\bm{x}$ with respect to
$\sigma$ to derive $\bm{x}_j^{\prime}= \bm{J}_j \left(\theta_j^{\prime},
\phi_j^{\prime} \sin\theta_j\right)^T$, where $\bm{J}_j$ is defined in
(\ref{map_3}). In (\ref{slow_dyn_a}) we then use the first expression in 
(\ref{alp_def}) for $\bm{\alpha}$ and pre-multiply both sides of
the resulting expression with $\bm{x}_j^{\prime}$. This yields that
\begin{equation*}
   \bm{x}_j^{\prime} = - \frac{2}{{\mathcal A}_j} \bm{J}_j \bm{J}_j^{T}
  \sum_{\substack{i=1 \\ i \neq j}}^N S_i \nabla_{\bm{x}} L_i 
 \bigr\rvert_{\bm{x}=\bm{x}_j} \,.
\end{equation*}
A direct calculation using (\ref{map_3}) shows that
$\bm{J}_j \bm{J}_j^T=\I-{\mathcal Q}_j$, where ${\mathcal Q}_j=\bm{x}_j\bm{x}_j^{T}$. In
addition, we have $\nabla_{\bm{x}} L_i \bigr\rvert_{\bm{x}=\bm{x}_j} =
{(\bm{x}_j-\bm{x}_i)/|\bm{x}_j-\bm{x}_i|^2}$. In this way, we get
\begin{equation}
   \bm{x}_j^{\prime} = - \frac{2}{{\mathcal A}_j} \left[ \I - {\mathcal Q}_j
\right] \sum_{\substack{i=1 \\ i \neq j}}^N S_i 
\frac{(\bm{x}_j-\bm{x}_i)}{|\bm{x}_j-\bm{x}_i|^2}\,. \label{on_sphere_0}
\end{equation}
We then multiply both sides of this expression by $\bm{x}_j^T$ to obtain
\begin{equation}
   \frac{1}{2} \dd{ |\bm{x}_j|^2}{\sigma} = \left(
  1- |{\bm x}_j|^2\right) C_j \,, \quad
   C_j \equiv -\frac{2}{{\mathcal A}_j} \sum_{\substack{i=1 \\ i \neq j}}^N \frac{S_i}
  {|\bm{x}_j-\bm{x}_i|^2} \left( |{\bm x}_j|^2 - |{\bm x}_j||{\bm x}_i|
  \cos\gamma_{ij}\right)\,, \label{sphere:dist}
\end{equation}
where $\gamma_{ij}$ is the angle between $\bm{x}_i$ and $\bm{x}_j$. If
$|\bm{x}_j(0)|=1$ for $j=1,\ldots,N$ and $\bm{x}_i(0)\neq \bm{x}_j(0)$
for $i\neq j$, then one solution to (\ref{sphere:dist}) is
$|\bm{x}_j(\sigma)|=1$ for all $\sigma\geq 0$, so that, as expected,
the centers of the spots remain on the unit sphere for all time. Along
this specific solution $C_j\neq 0$ since $S_i>0$ for $i=1,\ldots,N$.
Finally, (\ref{slow_cart}) follows from (\ref{on_sphere_0}) by noting
that $\left(\I-{\mathcal Q}_j\right)\bm{x}_j=0$ when
$\bm{x}_j^T\bm{x}_j=1$.

\section{Quasi-equilibrium spot patterns: existence and stability} \label{sec:qe_instab}

As we characterized in Principal Result \ref{thm:dyn},
quasi-equilibrium spot patterns will exhibit slow spot dynamics on a
long ${\mathcal O}(\epsilon^{-2})$ time-scale. However, as was shown
in \cite{rozada2014}, such patterns can be unstable on an ${\mathcal
  O}(1)$ time-scale in certain parameter regimes. In order to analyze
the stability of the quasi-equilibrium spot patterns, we must first
analyze the bifurcation behavior of the solution set of the nonlinear
algebraic system (\ref{constraint}) for the spot strengths
$S_1,\ldots,S_N$ for a given spatial configuration
$\lbrace{\bm{x}_1,\ldots,\bm{x}_N\rbrace}$ of spots. The stability
analysis in \cite{rozada2014} focused largely on quasi-equilibrium
spatial patterns for which the spots have a common spot strength. Our
goal here is to extend this prior analysis by identifying
  solutions to (\ref{constraint}) where the spots can have rather
different spot strengths. The stability of these patterns is analyzed
through an extension of the stability analysis of \cite{rozada2014}.
Our analysis below will consider the two asymptotic ranges $\E=\Oh(1)$
and $\E=\Oh(\nu^{1/2})$, where different behavior occurs.  Before
considering these ranges of $\E$, we first outline the stability
analysis of \cite{rozada2014}.

\subsection{Stability criterion for the quasi-equilibrium spot patterns} \label{sec:qe:stabcrit}

The stability analysis in \cite{rozada2014} allowed for perturbations
of the quasi-equilibrium spot pattern that are either radially
symmetric or non-radially symmetric in an ${\mathcal O}(\epsilon)$
neighborhood of each spot.

The linear stability of the quasi-equilibrium pattern with respect to
non-radially symmetric perturbations near each spot was studied in
\S{3.1} of \cite{rozada2014} from the numerical computation of an
eigenvalue problem. There, it was found that a spot centered at
$\bm{x}_j$ is unstable to a peanut-shape perturbation when $S_j >
\Sigma_2(f)$. The subscript on $\Sigma$ refers to instability with
respect to the local peanut-splitting angular mode $\cos{2\omega}$
where $\omega=\arg(\bm{x}-\bm{x}_j)$ as $\bm{x}\to\bm{x}_j$. The curve
$\Sigma_2$ versus $f$ is plotted in Fig.~4 of \cite{rozada2014}, and
we have
\begin{equation} \label{self_rep}
\Sigma_2(0.3) \approx 11.89\,, \quad \Sigma_2(0.4) \approx 8.21 \,, \quad
\Sigma_2(0.5) \approx 5.96\,, \quad \Sigma_2(0.6) \approx 4.41 \,,\quad
\Sigma_2(0.7) \approx 3.23 \,.
\end{equation}
This peanut-shaped unstable mode was found numerically in
\cite{rozada2014} to trigger, on an ${\mathcal O}(1)$ time-scale, a
nonlinear spot self-replication event for the $j^\textrm{th}$ spot
when $S_j>\Sigma_2(f)$.

In contrast to the non-radially symmetric case, the stability analysis
of the quasi-equilibrium spot pattern with respect to radially
symmetric perturbations near each spot is more intricate since this
analysis is based on properties of a globally coupled eigenvalue
problem (GCEP) (cf.~\cite{rozada2014}).  To formulate the stability
problem, we first linearize (\ref{full_all}) around the quasi-equilibrium
solution $u_\textrm{qe}$ and $v_\textrm{qe}$ by introducing $\psi$ and $N$ by
\begin{equation*}
    u= u_\textrm{qe} + e^{\lambda t} \psi \,, \qquad  
    v= v_\textrm{qe} + e^{\lambda t} N \,.
\end{equation*}
The spectral problem for $\psi$ and $N$ is singularly perturbed, with an
inner region near each spot and an outer region away from the spot locations.
We now summarize the singular perturbation analysis of \S{3.2--3.4} of
\cite{rozada2014}, for the formulation of the GCEP.

In terms of the core solution $V_{j0}$ and $U_{j0}$, the inner problem
near the $j^\textrm{th}$ spot is to determine the radially symmetric solution to
\begin{subequations}\label{eqn:psi_N}
\begin{equation} \label{eqn:bruss_m0}
  \dell_\rho \psi_j -
  \psi_j+2f U_{j0} V_{j0} \psi_j + f U_{j0}^2 N_j =
  \lambda \psi_j\,, \qquad \dell_\rho N_j +
  \psi_j - 2 U_{j0} V_{j0} \psi_j - U_j^2 N_j =
  0\,,
\end{equation}
subject to the boundary conditions
\begin{equation}
{\psi}_j^{\prime}(0) = {N}_j^{\prime}(0) = 0\,; \qquad
  \psi_j\to 0 \,, \quad {N}_j\sim\log\rho +
  B_{j} + o(1), \text{ as $\rho\to\infty$}\,, \label{eqn:bruss_m0_bc}  
\end{equation}
\end{subequations}
for $\psi_j(\rho)$, $N_j(\rho)$ on $0<\rho<\infty$, where $\dell_\rho
\equiv \partial_{\rho\rho} + \rho^{-1} \partial_\rho$. The key
quantity to calculate from the solution to this problem is
$B_j=B_j(S_j,\lambda)$ at each $f>0$.

The analysis in the outer region involves the eigenvalue-dependent
Green's function $G_{\lambda}(\bm{x};\bm{x}_j)$ on the sphere, defined for
$\lambda\neq 0$ by
\begin{equation}\label{G_lambda}
\dell_S G_{\lambda} -\tau \lambda G_{\lambda}= - \delta(\bm{x} - \bm{x}_j)\,,
\quad G_{\lambda} \sim -\frac{1}{2\pi} 
\log |\bm{x}-\bm{x}_j| + R_\lambda + o(1) \text{ as $\bm{x} \to 
\bm{x}_j$} \,,
\end{equation}
where $R_{\lambda}$ is independent of $\bm{x}_j$.  In terms of $G_\lambda$,
$R_\lambda$, and $B_j$, we then define a symmetric Green's matrix $\G_{\lambda}$
and a diagonal matrix $\B$ by
\begin{equation} \label{eqn:glambda}
 \G_{\lambda} \equiv \begin{pmatrix} R_{\lambda}& &G_{\lambda ij}\\ &
   \ddots \\ G_{\lambda ij} & & R_{\lambda} \end{pmatrix}\,, \qquad
 \B \equiv \begin{pmatrix} {B}_{1} & & 0\\ & \ddots \\ 0 & &
   {B}_{N} \end{pmatrix} \,,
\end{equation}
where $G_{\lambda ij}\equiv G_{\lambda}(\bm{x}_i;\bm{x}_j)$. In terms
of $\G_\lambda$ and $\B$, we then define the matrix
$\M=\M(\bm{S},\lambda,\tau,f)$ by
\begin{equation}
	\M \equiv \I+2\pi\nu\G_{\lambda} + \nu \B \,,
	\label{M}
\end{equation}
where $\nu=-{1/\log\epsilon}$ and $\I$ is the $N\times N$ identity
matrix.  In terms of $\M$, the following stability criterion was
derived in \S{3.4} of \cite{rozada2014}:

\begin{result}[Globally Coupled Eigenvalue Problem (GCEP)] \label{thm:stab} 
For $\epsilon\to 0$, the quasi-equilibrium pattern is unstable to locally
radially symmetric perturbations near each spot when
\begin{equation}
    \mbox{det}(\M)=0 \,, \label{criterion}
\end{equation}
for some $\lambda$ on the range $\mbox{Re}(\lambda)>0$. Alternatively,
the quasi-equilibrium pattern is linearly stable if
$\mbox{det}(\M)\neq 0$ for any $\lambda$ in $\mbox{Re}(\lambda)>0$.
\end{result}

The condition for a zero eigenvalue crossing was obtained as a
special case in \cite{rozada2014}. Here we derive this condition by
studying the singular limit for ${\cal G}_\lambda$ as $\lambda\to
0$. Since $G_\lambda\sim \left[4\pi\tau\lambda\right]^{-1}+ G$ as
$\lambda\to 0$, where $G$ satisfies (\ref{Gexact}), we 
obtain in terms of $\G$ and ${\mathcal E}_0$ of Principal Result 
\ref{thm:quasi} that
\begin{equation}
 2\pi \nu {\mathcal G}_\lambda\sim \mu {\mathcal E}_0 - \nu \G\,, \qquad
   \mu\equiv \frac{N \nu}{2\tau\lambda} 
  \left[ \tau \lambda (\log{4}-1) +1\right] \,, \label{mu_def}
\end{equation}
Since ${\mathcal E}_0$ has rank one, we can substitute this expression into
(\ref{M}) and then use the Sherman-Woodbury-Morrison formula to get for
$|\lambda|\ll 1$ that
\begin{equation}
\M \sim \left(\I +\mu{\mathcal E}_0\right)\left[\I -\nu (\I+\mu
  {\mathcal E}_0)^{-1} \left(\G - \B \right) \right] 
  \sim \left(\I
+\mu{\mathcal E}_0\right)\left[\I -\nu \left( \I
  -\frac{\mu}{1+\mu}{\mathcal E}_0\right)\left(\G-\B\right) \right] \,.
\end{equation}
Since the spectrum of $\I+\mu {\mathcal E}_0$ is known, we have for
$|\lambda|\ll 1$ that
\begin{equation}
   \mbox{det}(\M)=(1+\mu) \mbox{det}(\M_0)  \,, \qquad
  \M_0 \equiv \left[\I -\nu \left( \I
  -\frac{\mu}{1+\mu}{\mathcal E}_0\right)\left(\G-\B\right) \right] \,.
  \label{m0}
\end{equation}
Since $\mu/(1+\mu)\to 1$ as $\lambda\to 0$, it follows that a zero-eigenvalue
crossing occurs when
\begin{equation}
   \mbox{det}\left[\I -\nu \left(\I -{\mathcal E}_0\right)\left(\G-\B\right) 
\right] =0 \,, \label{m0:zero}
\end{equation}
where $\B$ is to be evaluated at $\lambda=0$. By
differentiating the core problem (\ref{U0V0eqn}) with respect to $S_j$
and comparing the resulting system with (\ref{eqn:psi_N}),
we conclude that the diagonal entries of $\B$ are
\begin{equation}
    \B_{j}(S_j,0) = \chi^{\prime}(S_j) \,. \label{bj:p0}
\end{equation}
The criterion (\ref{m0:zero}) for a zero eigenvalue crossing with
$(\B)_{jj}=\chi^{\prime}(S_j)$ was previously derived in
\cite{rozada2014}.  For $\lambda\ll 1$, our new criterion
$\mbox{det}(\M_0)=0$ in (\ref{m0}) will be used below to determine the
behavior of any eigenvalues of the GCEP near a zero eigenvalue
crossing.

The stability analysis below relies on determining the asymptotics of
$B_{j}(S_j,\lambda)$ as $S_j\to 0$.  The following new
result, proved in \ref{proof:bj}, gives the leading-order term in
$B_j$ as $S_j\to 0$ for any $\lambda$:

\begin{lemma}[Diagonal entries of $\B$] \label{lemma:bj} For $S_j\to 0$, 
we have from (\ref{eqn:psi_N}) that
\begin{subequations}\label{hatb1}
\begin{equation}
   B_j \sim -\frac{\hat{B}_0}{S_j^2} + {\mathcal O}(1) \,, \qquad
   \hat{B}_0 \equiv \frac{(1-f)d_0 (\lambda+1)}{\lambda+1-f}
   \frac{b}{2\K(\lambda)}\,, \label{hatb1_1}
\end{equation}
where $b\equiv \int_{0}^{\infty}\rho w^2\, \de{\rho}\approx 4.934$ and
${\cal K}(\lambda)$ is defined in terms of the unique solution
$w(\rho)>0$ of $\dell_\rho w - w + w^2=0$, with $w\to 0$ as $\rho\to
\infty$, by
\begin{equation}
  \K(\lambda) \equiv \int_{0}^{\infty} \rho w
  \left(L_0-\lambda\right)^{-1} w^2 \, \de{\rho} - \frac{b}{2} \,. \label{hatb1_2}
\end{equation}
\end{subequations}
Here $L_0$ is the local operator defined by $L_0\Phi\equiv \dell_\rho
\Phi - \Phi + 2w\Phi$. For $\lambda$ real, the function $\K(\lambda)$
satisfies
\begin{subequations}\label{k:prop}
\begin{equation}
   \K(0)={b/2} \,, \qquad \K^{\prime}(\lambda)>0 \text{ on $0<\lambda<\sigma_0$}\,, \qquad \K(\lambda)\to +\infty 
  \text{ as $\lambda \to \sigma_0^{-}$}\,. \label{k:prop_1}
\end{equation}
Here $\sigma_0>0$ is the unique positive eigenvalue with eigenfunction
$\Phi_0>0$ of $L_0\Phi=\sigma\Phi$, normalized as $\int_{0}^{\infty}
\rho \Phi_0^2 \, \de{\rho}=1$. For $\lambda=\sigma_0-\delta$ with
$\delta\to 0^{+}$, we have
\begin{equation}
    \K(\lambda) \sim {C/\delta} + {\mathcal O}(1) \,, \qquad
   C \equiv \biggl(\int_{0}^{\infty} \rho w^2 \Phi_0 \, \de{\rho} \biggr)
  \biggl(\int_{0}^{\infty} \rho w \Phi_0 \, \de{\rho} \biggr)\,. \label{k:prop_2}
\end{equation}
For $\lambda=0$, and with $d_0$ and $d_1$ as defined in (\ref{chi:small_S}),
we have the two-term expansion
\begin{equation}
    B_{j}(S_j,0) =\chi^{\prime}(S_j) \sim  -\frac{d_0}{S_{j}^2} + d_1 \,, 
  \quad \mbox{as} \quad S_j \to 0 \,. \label{bj:p1}
\end{equation}
\end{subequations}
\end{lemma}

\subsection{An overview of the quasi-equilibria solution} \label{sec:qe_overview}

Before deriving the asymptotic form of the spot strengths, we first
explore the global bifurcation structure and solve the full nonlinear
algebraic system \eqref{constraint} for a particular arrangement of $N
= 2$ spots. Numerical solutions of the system for different values of
$\E$ and $\nu$ are found using the continuation and bifurcation
software AUTO-07P, and the continuation process is initiated by using,
as an initial guess, the results from the $\nu \to 0$ asymptotics (to
be derived in the next section).

\begin{figure}[htb] \centering
\includegraphics[scale=1.1]{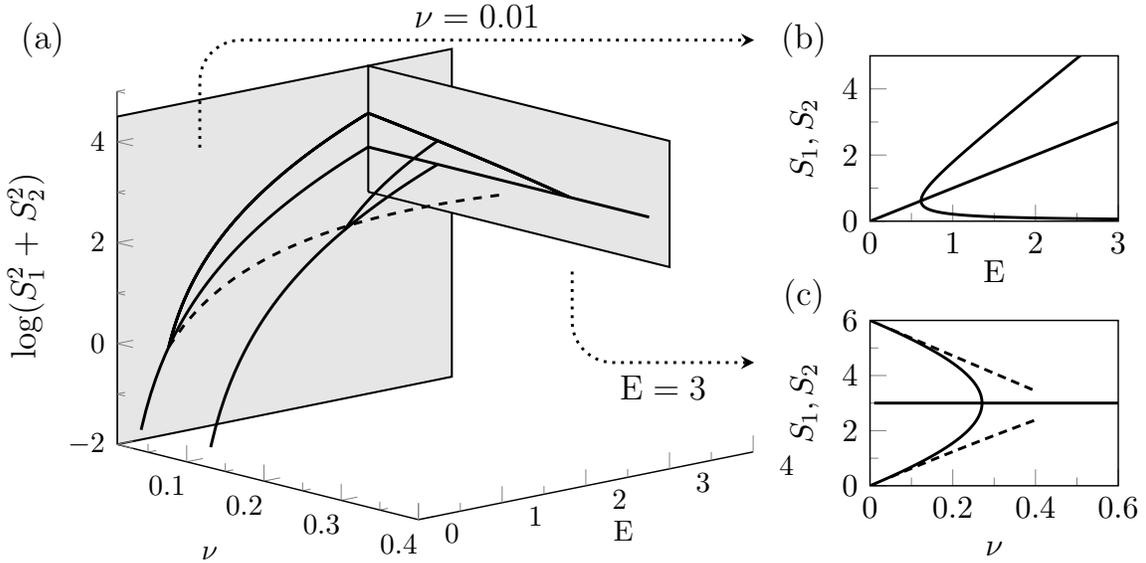}
\caption{Bifurcation diagrams resulting from numerical solutions of
  \eqref{constraint} with $f = 0.3$ for the case of $N = 2$ spots
  centered at $(\phi,\theta)=(0,{\pi/2})$ and
  $(\phi,\theta)=(\pi,{\pi/2})$. The larger plot (a) shows the
  three-dimensional bifurcation structure, with $\log(S_1^2 + S_2^2)$
  as a function of $\nu$ and $\E$. The dashed path corresponds to $\E
  = \sqrt{d_0\nu}$ where $S_{1,2} = \sqrt{d_0}\nu$ from
  \eqref{commonE}. Individual spot strengths corresponding to fixed
  values of $\E$ and $\nu$ are shown in (b) and (c). The dashed line
  (c) corresponds to Type II asymmetric solutions, given from
   \eqref{type2}. To leading order they are $S_1 \sim
  2\E$ and $S_2 \sim {\nu d_0/(2\E)}$. In (b), the solution
  $S_1$ and $S_2$ to the reduced system (\ref{lo:small_E}), valid for
  $\E=\Oh(\nu^{1/2})$, overlays almost exactly with the full numerical
  solution. \label{fig:auto_N2}}
\end{figure}

First, examine Figure~\ref{fig:auto_N2}(a), which corresponds to the
case of $f = 0.3$ and $N = 2$ spots centred at $(\phi, \theta) = (0,
\pi/2)$ and $(\phi, \theta) = (\pi, \pi/2)$. The numerically computed
bifurcation structure resides within $(\nu, \E, \log ||\bm{S}||_2^2)$
space, and for each point in the $(\nu, \E)$ plane, there is either
one or two possible quasi-equilibria, distinguished by the size of the
norm $\|\bm{S}\|_2^2$.

For a fixed value of $\E$, in subfigure \ref{fig:auto_N2}(c) we plot
the curves $S_{1,2}$ versus $\nu$. Note that when $N = 2$, the
matrix $\G$ is cyclic, and there exists a solution of
\eqref{constraint} with the common spot strength, $S_c = 2\E/N$. This
corresponds to the flat line $S_{1,2} = 3$ in the subfigure. We shall
call these Type I patterns. We also see that when $\nu$ is
sufficiently small, there appears to be an additional asymmetric
pattern that bifurcates from the Type I branch, with one small
$\Oh(\nu)$ spot and one large $\Oh(1)$ spot. We refer to these as Type
II patterns. Both Type I and II solutions are studied in
\S\ref{qe:big}.

However, it is apparent from Figure~\ref{fig:auto_N2} that there also
exists a distinguished limit if $\E \to 0$ simultaneously as $\nu \to
0$. This is shown via the curves $(\E, S_{1,2})$ in subplot (b). As
similar to Type I and II solutions, there is a shared curve where $S_1
= S_2$ (the centre curve of the subplot), and two flanking curves
corresponding to a small and large spot, which bifurcate from the
centre branch. In \S\ref{qe:big}, we demonstrate that the
distinguished limit is described by $\E = \Oh(\sqrt{\nu})$ as $\nu \to
0$, and the solutions shown in (b) near the bifurcation point
correspond to spot strengths of $\Oh(\sqrt{\nu})$. We call these Type
III patterns and they correspond to both equal and unequal spot
strengths. We will also later derive the formula for the dashed line,
$\E = \E(\nu)$ in Figure~\ref{fig:auto_N2}(a), which describes the
critical bifurcation point of the $\E, \nu \to 0$ limit, where the
asymmetric branches split from symmetric branch.

For $N > 2$, the situation is more complex in the
case of the asymmetric Type II patterns, and there may be $m<N$ spots
of strength $\Oh(1)$ and $(N - m)$ spots of strength
$\Oh(\nu)$. However, the classification remains the same, and we can
expect the following three types of solutions:

\begin{equation}
\begin{alignedat}{3}
\textrm{Type I (symmetric):} \qquad & S_j = \Oh(1),& \qquad &j = 1, 2, \ldots, N, \\
\textrm{Type II (asymmetric):} \qquad & S_j = \Bigg\lbrace
\begin{aligned}
&\Oh(1),\\
&\Oh(\nu),  
\end{aligned} &
&\!\begin{aligned}
 j &= 1, 2, \ldots, m, \\
 j &= m+1, \ldots, N,
\end{aligned}\\
\textrm{Type III (a/symmetric):} \qquad & S_j = \Oh(\nu^{1/2}),& \quad &j = 1, 2, \ldots, N.
\end{alignedat}
\end{equation}

We now comment on the splitting of the asymmetric branches from the
symmetric branches for general number of spots. If the spot locations,
$\bm{x}_j$, for $j = 1, \ldots, N$ are distributed in such a way that
\begin{equation} \label{Ge}
  \G \bm{e} = k_1 \bm{e} \,,
\end{equation}
then a solution to (\ref{constraint}) is the equal spot-strength solution
$\bm{S} = S_c \bm{e}$ where 
\begin{equation} \label{Sc}
  S_c = \frac{2\E}{N} \,.
\end{equation}
The property (\ref{Ge}) holds for any two-spot pattern, for a pattern
of equally spaced spots on a ring of constant latitude, for spots
centered at the vertices of any of the platonic solids (see
Table 1 of \cite{rozada2014} and \S\ref{sec:dynamics} below).

Assuming that $N>1$ and that (\ref{Ge}) holds, then a bifurcation
occurs if and only if the Jacobian matrix of $\N(\bm{S})$ in
\eqref{constraint} is singular when $\bm{S}=S_c\bm{e}$. By setting
$\bm{S} = S_c\bm{e} + \bm{\Phi}$, with $|\bm{\Phi}| \ll 1$ in
\eqref{constraint}, a bifurcation from the symmetric solution branch
occurs if and only if there exists a non-trivial $\bm{\Phi}$ to
\begin{equation} \label{qs2}
\Bigl[ \I - \nu (\I - {\mathcal E}_0) (\mathcal{G}-\chi^{\prime}(S_c)\I)
 \Bigr] \bm{\Phi} = 0 \,.
\end{equation}
Upon comparing (\ref{qs2}) with (\ref{m0:zero}), we observe that this
bifurcation point corresponds to a zero-eigenvalue crossing, and hence
an exchange of stability for the symmetric solution branch. Since $\G$
is a symmetric matrix with $\G\bm{e} = k_1 \bm{e}$, it follows that
there exists eigenvectors, $\bm{q}_j$, with $\G\bm{q}_j = k_j
\bm{q}_j$, for $j = 2, \ldots, N$, where $\bm{q}_j^T \bm{e}=0$. It is
readily verified that $\bm{\Phi}=\bm{q}_j$ satisfies (\ref{qs2}) when
$S_c = S_{c_j}$ for $j=2,\ldots,N$, where $S_{c_j}$ satisfies the
nonlinear algebraic equation
\begin{equation} \label{Scj}
 \nu^{-1} - k_j + \chi^{\prime}(S_{c_j}) = 0 \,, \qquad j=2,\ldots,N\,.
\end{equation}
From \eqref{Sc}, this indicates that a bifurcation occurs at the $j =
2, \ldots, N$ points where
\begin{equation} \label{Ejj}
\E = \E_j = \frac{NS_{c_j}}{2}.
\end{equation}
For $\nu\ll 1$, this yields $\E_j={\mathcal O}(\nu^{1/2})$.

Note, however, that it may be the case that the eigenvalues,
$\bm{q}_j$, are not distinct, and in particular, this can certainly
occur if, \emph{e.g.} the spots are arranged on a plane of constant
latitude and $\mathcal{G}$ is a cyclic matrix. In this case, the
number of bifurcating branches will still be $N - 1$, but the number
of bifurcation points (in $\E$) will be equal to the number of
distinct eigenvalues. In \S\ref{qe:small}, we will derive the
dashed curve $\E = \E(\nu)$ shown in Figure~\ref{fig:auto_N2}(a),
which is a case of \eqref{Ejj} in the uniform limit of $\E \to 0$ and
$\nu \to 0$, and where the bifurcation points coalesce.

\subsection{Quasi-equilibria for $\E=\Oh(1)$ (Type I and II)} \label{qe:big}


We first consider the symmetric Type I patterns, for which all spots
are characterized by $S_j = \Oh(1)$. For $\nu={-1/\log\epsilon}\ll 1$,
a two-term regular perturbation expansion of (\ref{constraint}) yields
that
\begin{equation}\label{reg:s}
   \bm{S} \sim \frac{2E}{N} \left[ \bm{e} +\nu (\I - {\mathcal E}_0 )
   {\mathcal G} \bm{e} + {\mathcal O}(\nu^2) \right]\,.
\end{equation}
Here $\bm{S}=(S_1,\ldots,S_N)^T$, $\bm{e}=(1,\ldots,1)^T$, ${\mathcal
  E}_0$ and ${\mathcal G}$ are defined in Principal Result
\ref{thm:quasi}. 

To determine the stability property of Type I patterns, we observe
from (\ref{M}) that ${\mathcal M}=I+{\mathcal O}(\nu)$ as $\nu\to 0$
when $\bm{S}={\mathcal O}(1)$ and $\lambda={\mathcal O}(1)$. In
addition, from (\ref{m0}), we have ${\mathcal M}_0=I+{\mathcal
  O}(\nu)$ for $\lambda=0$. As such, since both $\M$ and $\M_0$ are
non-singular for $\nu\to 0$ when $\bm{S}={\mathcal O}(1)$, we conclude
from the GCEP criterion in Principal Result \ref{thm:stab} that this
class of spot pattern is linearly stable to radially symmetric
perturbations near each spot when $\nu\ll 1$. As such, the stability
criterion for this class of solutions is simply that $S_j<\Sigma_2(f)$
to prevent spot self-replication instabilities triggered by a locally
non-radially symmetric perturbation near the $j^\textrm{th}$ spot.

Next, consider Type II patterns. Suppose that there are $m\geq 1$
small spots, with $S_j={\mathcal O}(\nu)$ for $j=1,\ldots,m$, and
$N-m$ large spots with $S_j={\mathcal O}(1)$ for $j=m+1,\ldots,N$. By
using $\chi(S)\sim {d_0/S}$ as $S\to 0$ in \eqref{chi:small_S}, a perturbation
calculation on (\ref{constraint}) shows that the spot-strengths for
this pattern have the following two-term asymptotics for $\nu\ll 1$:
\begin{subequations}\label{type2}
\begin{equation} \label{type2:exp}
   S_j \sim \begin{cases}
   S_{0}^{\star} + \nu S_{j1}^{\star} + \cdots & \text{for $j = m + 1, \ldots , N$} \\
   \nu S_{0} + \nu^2 S_{j1} + \cdots & \text{for $j=1,\ldots,m$} 
   \end{cases}\,,
\end{equation}
where $S_0$, $S_{j1}$, $S_{0}^{\star}$, and $S_{j1}^{\star}$, are given by
\begin{alignat}{3} \label{type2:coeff}
  S_0^{\star} &= \frac{2\E}{N-m} \,, 
      & \quad S_{j1}^{\star} 
      &= -\frac{md_0}{2\E} + \frac{2\E}{N-m} \Lsum_j \,,  \\
  S_0 &= \frac{d_0(N-m)}{2\E} \,, 
      & \quad S_{j1} 
      &= \frac{d_0(N-m)^2}{8\E^3} \left[d_0N -
  2\E \chi(S_{0}^{\star}) - \frac{4\E^2}{(N-m)} \Lsum_j\right] \,.
 \end{alignat}
Here $d_0={b(1-f)/f^2}$ from (\ref{chi:small_S}), while $\Lsum_j$ is 
defined by
\begin{equation}
  \Lsum_j \equiv \sum_{\substack{i=m+1 \\ i \neq j}}^N L_{ij} - \frac{1}{N-m}
  \sum_{\substack{i=m+1 \\ i \neq k}}^N   \sum_{k=m+1}^N  L_{ik} \,, \qquad
   L_{ij}\equiv \log|\bm{x}_i-\bm{x}_j| \,. \label{type2:sum}
\end{equation}
\end{subequations}
From the criterion in Principal Result \ref{thm:stab}, we now
show that these Type II patterns are all unstable.

\begin{result}[Stability of Type II patterns] \label{thm:mixed} 
For $\epsilon\to 0$, the Type II quasi-equilibrium patterns with
spot strengths in (\ref{type2}) are all unstable on an ${\mathcal
  O}(1)$ time-scale.
\end{result}

\begin{proof}
For $\nu\ll 1$, we show that $\mbox{det}(\M)=0$ for some $\lambda$ on
the positive real axis that is ${\mathcal O}(\nu)$ close to the
eigenvalue $\sigma_0>0$ of the local operator $L_0$ defined in Lemma
\ref{lemma:bj}.  We set $\lambda=\sigma_0-\delta_0\nu$ for some
$\delta_0>0$, and look for a root of (\ref{criterion}) where $\M$ is
defined in (\ref{M}). From (\ref{hatb1_1}) and (\ref{k:prop_2}) of Lemma
\ref{lemma:bj}, and (\ref{type2}), we obtain for the small spots that
$B_j={\mathcal O}(\nu^{-1})$, with
\begin{equation}
  B_{j} \sim - \frac{4\E^2}{\nu d_0^2(N-m)^2} \delta_0 \hat{B}_{0-} \,, \qquad
  \hat{B}_{0-} \equiv  \frac{(1-f)d_0 (\sigma_0+1)}{\sigma_0+1-f}
   \frac{b}{2C}\,, \quad j=1,\ldots,m \,,  \label{bj_small}
\end{equation}
where $C>0$ is defined in (\ref{k:prop_2}). In contrast, for the large
spots we have $B_j={\mathcal O}(1)$ for $j=m+1,\ldots,N$. Upon
substituting (\ref{bj_small}) into (\ref{M}), we obtain that
\begin{equation}
   \M = \I -  \frac{4\E^2}{d_0^2(N-m)^2} \delta_0 \hat{B}_{0-}
\begin{pmatrix} \I_m & 0 \\ 
                 0   & 0 
 \end{pmatrix} + {\mathcal O}(\nu)  \,,  \label{m_small}
\end{equation}
where $\I_m$ is the $m\times m$ identity matrix. Upon setting $\det(\M)=0$, 
we get that $\M$ is singular when
\begin{equation}
   \delta_0 = \frac{d_0^2(N-m)^2}{4\E^2 \hat{B}_{0-}} = \frac{d_0 (N-m)^2}
 {2\E^2} \frac{ (\sigma_0+1-f) C }{(1-f)(\sigma_0+1)b} >0  \,. \label{delta_0}
\end{equation}
Thus, for Type II patterns the GCEP has an eigenvalue 
$\mbox{Re}(\lambda)>0$ with asymptotics $\lambda=\sigma_0-{\mathcal O}(\nu)$.
\end{proof}

\subsection{Quasi-Equilibria for $\E={\mathcal O}(\sqrt{\nu})$ (Type III patterns)}\label{qe:small}

As shown in Fig.~\ref{fig:auto_N2}, there exists a distinguished limit
when both $\E$ and $\nu \to 0$ simultaneously, leading to Type III
patterns. The correct scaling that captures this limit is $\E =
\Oh(\sqrt{\nu})$ and we introduce the re-scaled new variables
$\tilde{S}_j$, $\tilde{E}$, and $\tilde{v}$, defined by
\begin{equation*}
  S_j = \tilde{S}_j \nu^{1/2}\,, \qquad
  \E = \tilde{\E} \nu^{1/2}\,, \qquad  
  v_c = \tilde{v} \nu^{1/2}\,,
\end{equation*}
into the alternative form (\ref{match_3a}) of the nonlinear 
system for the spot strengths. Upon using $\chi(S_j)\sim {d_0/S_j}$ as
$S_j\to 0$ from \eqref{chi:small_S}, we obtain that $\tilde{S}_j$ for
$j=1,\ldots,N$ and $\tilde{v}$ satisfy the leading-order result
\begin{equation} \label{quadeq}
 {\mathcal H}(S_j) \equiv \tilde{S}_j + \frac{d_0}{\tilde{S}_j} = \tilde{v}
 \,, \qquad \sum_{j=1}^{N} \tilde{S}_j = 2\tilde{\E} \,,
\end{equation}
where $d_0$ is given in (\ref{chi:small_S}). The function ${\mathcal
  H}(\xi)$ in (\ref{quadeq}) is convex for $\xi>0$ and satisfies
${\mathcal H}(\xi)\to +\infty$ as $\xi\to 0^{+}$ and as $\xi\to
\infty$. It has a global minimum at $\xi=\sqrt{d_0}$ with minimum
value ${\mathcal H}(\sqrt{d_0})=2\sqrt{d_0}$.

\begin{figure}[htb]\centering
\includegraphics{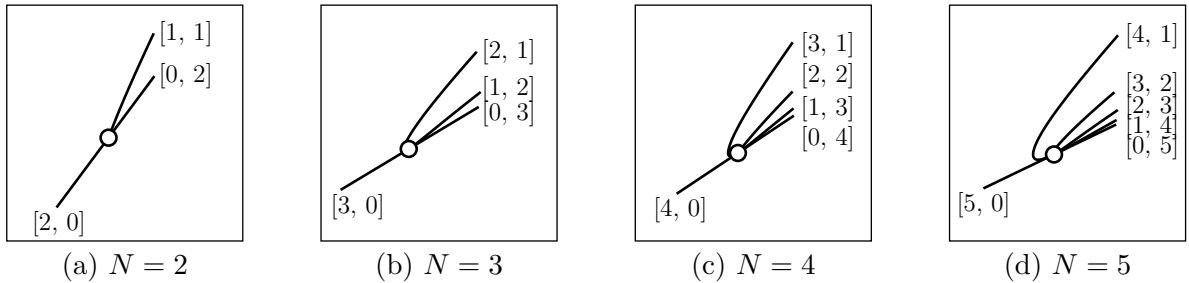}
\caption{Bifurcation diagrams for the leading-order problem
  (\ref{lo:small_E}). The horizontal axis corresponds to $\tilde{\E}$ and the vertical axis to the solution measure $(\tilde{S}_1^2 + \ldots \tilde{S}_N^2)$. The circular nodes correspond to where asymmetric branches bifurcate from the symmetric solution branch.
  The notation corresponds to $[N_-, N_+]$, the number of 
  $\tilde{S}_-$ and $\tilde{S}_+$ spots. \label{minibif}}
\end{figure}

With these properties of ${\mathcal H}(\xi)$, it follows that each
spot can either be of small spot strength, $\tilde{S}_{-}$, or large spot
strength, $\tilde{S}_+$, where $0<\tilde{S}_{-}\leq \sqrt{d_0} \leq
\tilde{S}_{+}$. To construct an asymmetric pattern with $N_-$ small
spots and $N_+=(N-N_-)$ large spots, we must solve the leading-order
problem
\begin{equation} \label{lo:small_E}
   {\mathcal H}(\tilde{S}_{-})= {\mathcal H}(\tilde{S}_{+}) \,, \qquad 
 N_-\tilde{S}_{-} + (N-N_-) \tilde{S}_{+}= 2\tilde{E} \,.
\end{equation}
The bifurcation point where asymmetric quasi-equilibria emerge from
the common spot strength solution branch is obtained by setting
$N_-=0$ and $\tilde{S}_{-}=\tilde{S}_{+}$, which yields
\begin{equation} \label{commonE}
\tilde{E} \sim \frac{N\sqrt{d_0}}{2} \quad \text{and} \quad
\tilde{S}_{-} = \tilde{S}_+ \sim \sqrt{d_0}\,.
\end{equation}
For different $N_-$ and $N_+$, in Fig.~\ref{minibif} we plot
$\sum_{j=1}^{N}\tilde{S}_j^2$ versus $\tilde{E}$, as computed from
(\ref{lo:small_E}), illustrating the symmetric and asymmetric solution
branches.

Notice furthermore that the asymmetric branches for (\ref{lo:small_E})
that emerge from the bifurcation point (with the symmetric branch) can
be continued into the regime where $\tilde{E}={\mathcal
  O}(\nu^{-1/2})$, or equivalently where $\E={\mathcal O}(1)$. These lead to the unstable Type II mixed patterns studied in \S\ref{qe:big}, which consist of both small and large spots. This is the connection between the two shaded planes in
Fig.~\ref{fig:auto_N2}.

However, the question of whether the prediction of a common
bifurcation point from this leading-order system (\ref{lo:small_E}) is
robust to perturbations in $\nu$ from the full system
\eqref{constraint} is another matter entirely, and is found to depend
on whether the condition (\ref{Ge}) on the Green's matrix holds or not
(see Fig.~\ref{fig:auto_N3}). When (\ref{Ge}) holds, (\ref{Scj}) will
be used below in (\ref{Ej}) to show that, for $N>2$, higher order in
$\nu$ terms lead to transcritical bifurcation points in $\E$ that are
${\mathcal O}(\nu^{3/2})$ close.

\subsection{Comparisons with numerical results}

The conclusion from our analysis in \S\ref{sec:qe_overview} and
from Fig.~\ref{fig:auto_N2} regarding the global bifurcation structure
for $N = 2$ is as follows. First, for $N = 2$, the common solution
with $\bm{S} = \E \bm{e}$ is an exact solution for all $\nu$ for any
two-spot pattern. This follows since $\G$ is cyclic for any two-spot
configuration. Second, in the limit $\nu \to 0$ with $\E =
{\mathcal O}(1)$, the Type II patterns are given by setting $m=1$ and
$N=2$ in (\ref{type2}). Third, for $\nu\to 0$ with $\E={\mathcal
  O}(\nu^{1/2})$ the asymmetric quasi-equilibrium is characterized by
(\ref{lo:small_E}), and indeed bifurcates from the symmetric solution
branch for any $\nu>0$ small. This bifurcation, calculated from
\eqref{commonE}, is shown in the dashed curve in Fig.~\ref{fig:auto_N2}(a).

Recall from \S\ref{sec:qe_overview} that whenever the Green's
matrix $\G$ satisfies \eqref{Ge} there is a solution (for all $\nu$)
where the spots have a common strength. Typically, there is a
degenerate eigenvalue for $\G$ of multiplicity two in the subspace
perpendicular to $\bm{e}$. This must necessarily be true if $\G$ is
cyclic.  

\begin{figure}[htb] \centering
\includegraphics{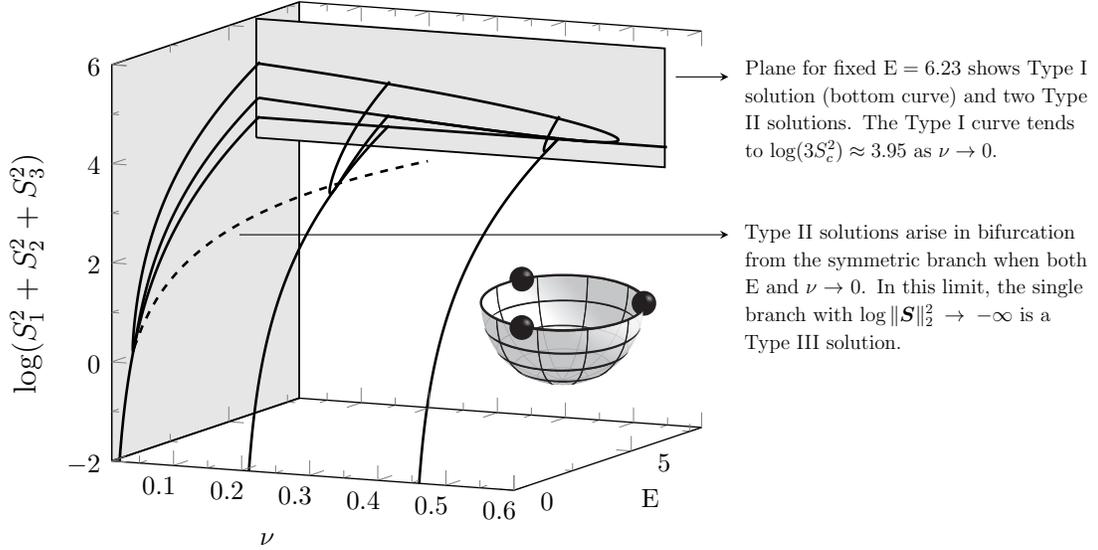} 
\caption[scale=1.1]{Bifurcation diagrams of \eqref{constraint} in $(\nu, \E,
  \|\bm{S}\|_2^2)$ space corresponding to $N = 3$ and $f = 0.3$. The
  vertical axis is $\log(S_1^2 + S_2^2 + S_3^2)$. The spots
  centered symmetrically at 
  $(\phi,\theta)=\{(0,\pi/2),(2\pi/3,\pi/2),(4\pi/3,\pi/2)\}$. For
  small values of $\nu$, there are two Type II patterns originating
  from a common bifurcation point from the symmetric solution branch
  in the $\E=\Oh(\nu^{1/2})$ regime. The
  two planes correspond to $\nu = 0.01$ and $\E =
  6.23$. \label{fig:auto_N3a}}
\end{figure}

\begin{figure}[htb] \centering
\includegraphics[scale=1.1]{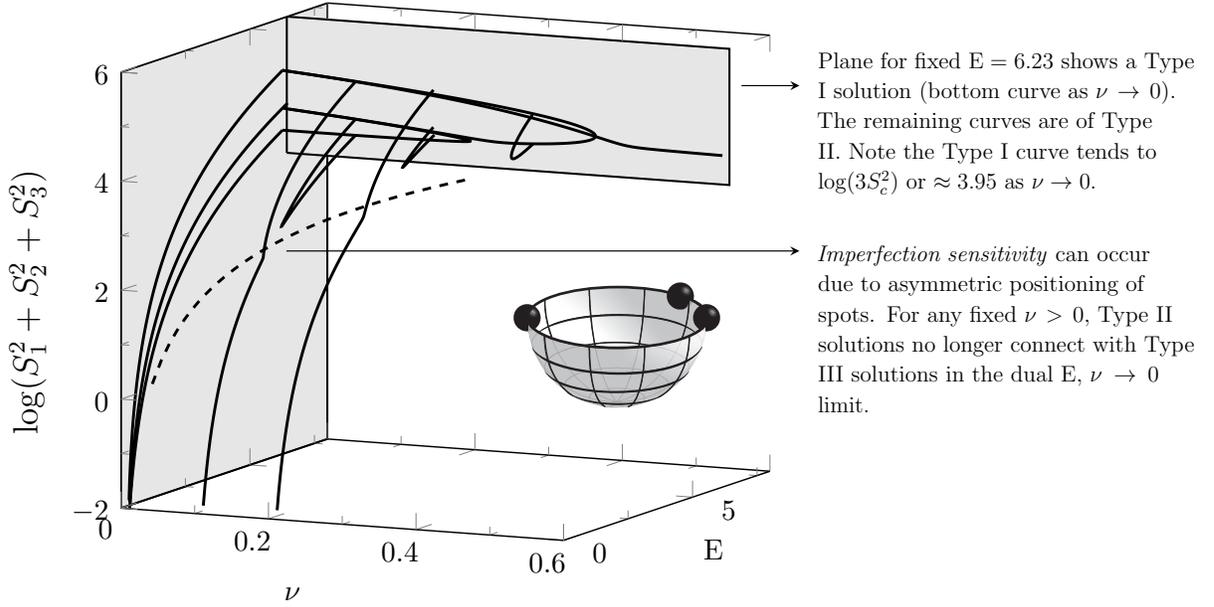} 
\caption{Same as for Fig.~\ref{fig:auto_N3a} but with spots are centered
  asymmetrically at $(\phi,\theta)=\{(0,{\pi/2}), ({\pi/4},{\pi/2}),
  (\pi,{\pi/2})\}$. For small values of $\nu$, there are two Type II
  patterns for $\E=\Oh(1)$ that do not originate from transcritical
  bifurcations in the $\E=\Oh(\nu^{1/2})$ regime. The
  two planes correspond to $\nu = 0.01$ and $\E =
  6.23$. \label{fig:auto_N3b}}
\end{figure}

We now consider the case $N=3$ and study the effect on the bifurcation
structure of solutions to (\ref{constraint}) on whether \eqref{Ge}
holds or not.  In Figs.~\ref{fig:auto_N3a} and \ref{fig:auto_N3b}, we show numerical solutions
for $f = 0.3$ and $N = 3$ spots of two different spatial
configurations. The results in Fig.~\ref{fig:auto_N3a} correspond to when the
spots are placed equidistantly along the equator, and \eqref{Ge}
holds, while for the other figure, the spots are placed
asymmetrically along the equator, so that \eqref{Ge} does not
hold. The bifurcation curves are plotted in $(\nu, \E, \log \| \bm{S}
\|^2_2)$ space.

In both configurations, when $\E = \Oh(1)$ is fixed, we observe two
Type II patterns in the $\nu \to 0$ limit. These solutions are found
by setting $(m,N)=(1,3)$ and $(m,N)=(2,3)$ in \eqref{type2}. For the
symmetric arrangement of Fig.~\ref{fig:auto_N3a}, $\bm{S}={2\E\bm{e}/3}$ is a
solution for all $\nu>0$, and for sufficiently small $\nu$, it is
observed that the Type II patterns bifurcate from the symmetric branch
in the $\E=\Oh(\nu^{1/2})$ regime at a common bifurcation point. In
the $\nu \to 0$ limit, the common bifurcation point is given by
\eqref{commonE}, and as seen in the figure, the agreement with the
numerical solutions is very good. For this case, $\G$ is a cyclic
matrix, so that there is only one eigenvalue of multiplicity two in the
subspace orthogonal to $\bm{e}$. As such, from (\ref{Scj}), there is
still a common bifurcation point when higher order terms in $\nu$
are included, and indeed this is evident from the figure.

However, for the asymmetric arrangement of Fig.~\ref{fig:auto_N3b}, where (\ref{Ge}) does not hold, we observe
that for any $\nu>0$ the Type II solution branch does not
undergo a transcritical bifurcation when path-followed into the
$\E=\Oh(\nu^{1/2})$ regime. This figure shows that the leading-order $\nu=0$
approximation (\ref{lo:small_E}), which predicts a common bifurcation
point, is not robust to perturbations in $\nu>0$ and, therefore, exhibits
{\em imperfection sensitivity} to higher order terms.

\subsection{Stability criterion for $\E = \Oh(\sqrt{\nu})$ (Type III patterns)}
\label{sec:type3}

We now return to the issue of stability discussed in
\S\ref{sec:qe:stabcrit}, but make use of the limit $\E, \nu \to 0$
derived in \S\ref{qe:small} in order to focus on the behaviour near
the critical bifurcation points $\E = \E_j$ given in \eqref{Ejj}.

By using (\ref{bj:p1}) for
$\chi^{\prime}(S_j)$ as $S_j\to 0$ in (\ref{Scj}), and then letting 
$\nu\to 0$, we obtain
\begin{equation}
  \E_j \sim \frac{N\sqrt{\nu d_0}}{2} \left[ 1 - \nu (d_1-\kappa_j)
 + {\mathcal O}(\nu^2) \right] \,, \qquad j=2,\ldots,N \,. \label{Ej}
\end{equation}

Again, we remark that the eigenvalues $k_j$ for $j=2,\ldots,N$ of $\G$ in the
subspace perpendicular to $\bm{e}$ are in general not distinct. This
eigenvalue degeneracy is necessarily the case when $\G$ is a cyclic
matrix. In this case, the number of bifurcating branches is $N-1$, but
the number of bifurcation points in $\E$ is the number of distinct
$k_j$ in $j=2,\ldots,N$.

From (\ref{Ej}), the leading-order stability threshold is $\E\sim
\E_c$ with $\E_c \equiv {N\sqrt{\nu d_0}/2}={\mathcal
  O}(\sqrt{\nu})$. To analyze the zero eigenvalue crossing as $\E$
crosses above $\E_c$, we use (\ref{m0}) together with
$\B\sim -\hat{B}_0 S_c^{-2}\I$ for $S_c={2\E/N}\ll 1$, to get for
$\lambda\ll 1$ that
\begin{equation*}
 \M_0 \bm{q}_j = \bm{q}_j - \nu \left(\I - \frac{\mu}{1+\mu} {\mathcal
   E}_0 \right) \left(\G - \B\right) \bm{q}_j = \left(1 - \nu \kappa_j
 + \nu\frac{\hat{B}_0}{S_c^2}\right) \bm{q}_j \,,
\end{equation*}
where $\G \bm{q}_j=\kappa_j \bm{q}_j$ for $j=2,\ldots,N$. Therefore,
$\mbox{det}(\M_0)=0$ for $|\lambda|\ll 1$ when
\begin{equation}
  \frac{1}{\nu} - \kappa_j + \frac{\hat{B}_0}{S_c^2} =0 \,, \label{zero:c}
\end{equation}
where $S_c={2\E/N}$ and $\hat{B}_0$ is defined in (\ref{hatb1_1}). By
solving (\ref{zero:c}) for $\E$, we obtain to leading order in $\nu$ that
\begin{equation}
   \left( \frac{\E}{\E_c}\right)^2 = \Z(\lambda) \,, 
  \qquad \Z(\lambda) \equiv (1-f) \left( \frac{\lambda +1}{\lambda + 1-f}
  \right) \frac{b}{2\K(\lambda)} \,. \label{z:cross}
\end{equation}
Upon using the properties of $\K(\lambda)$ in (\ref{k:prop_1})
we conclude that $\Z(0)=1$, and we calculate
\begin{equation}
  \Z^{\prime}(\lambda) = \frac{(1-f)b}{2} \left( 
-\frac{f}{\K(\lambda) (\lambda+1-f)^2} - \frac{(\lambda+1)}{(\lambda + 1 -f)}
   \frac{\K^{\prime}(\lambda)}{\left[\K(\lambda)\right]^2} \right) < 0
\end{equation}
on $0<\lambda<\sigma_0$. Therefore, for any $\E<\E_c$ with $\E-\E_c$ small, 
there exists a unique $\lambda^{\star}\ll 1$ with $\lambda^{\star}>0$.

We conclude that the zero eigenvalue crossing is such that the
symmetric solution branch is unstable for $\E<\E_c = {N\sqrt{\nu
    d_0}/2}$ for $\E-\E_c$ small. For $\E>\E_c$ with $E-\E_c$ small,
the spectrum of the linearization around the symmetric solution has no
unstable real eigenvalues. Through the detailed analysis of a nonlocal
eigenvalue problem, it was shown in \S 4.4 of \cite{rozada2014} that
in fact there are no unstable eigenvalues in $\mbox{Re}(\lambda)>0$,
and consequently the symmetric solution branch is linearly stable when
$\E>\E_c$ with $\E={\mathcal O}(\sqrt{\nu})$.

\section{A selection of results for spot dynamics}\label{sec:dynamics}

In this section we give some results for spot dynamics as obtained by
solving the DAE system (\ref{slow_cart}) and (\ref{constraint})
numerically with $\E={\mathcal O}(1)$. Based on the stability analysis
of \S\ref{sec:qe_instab}, we only consider patterns for which
$S_j={\mathcal O}(1)$ as $\nu\to 0$. The slow dynamics
(\ref{slow_cart}) is valid provided that each $S_j$ is below the spot
self-replication threshold, i.e.~$S_j<\Sigma_2$ for $j=1,\ldots,N$.
For a two-spot pattern the following result, as proved in
\ref{proof:two-spot}, provides an explicit solution to the DAE system:
\begin{lemma}[Explicit two-spot solution]\label{lemma:two-spot} 
Let $\gamma_{1,2}=\gamma_{1,2}(\sigma)$ 
denote the angle between the spot centers $\bm{x}_1$ and
$\bm{x}_2$, i.e. $\bm{x}_2^T\bm{x}_1=\cos\gamma_{1,2}$.  Then, provided
that $E<\Sigma_{2}(f)$, we have for all time $\sigma=\epsilon^2 t\geq 0$ that
\begin{equation}
   \cos\left( {\gamma_{1,2}/2}\right) = \cos\left( {\gamma_{1,2}(0)/2}\right) 
 e^{-{\E \sigma/\vert {\mathcal A}(\E)\vert}} \,. \label{2spot}
\end{equation}
Since $\gamma_{1,2}\to \pi$ as $\sigma\to \infty$ for any
$\gamma_{1,2}(0)$, the steady-state two-spot
pattern will have spots centered at antipodal points on the sphere for any
initial configuration of spots.
\end{lemma}

Before proceeding, we also note that in in \eqref{slow_dyn_e} and
\eqref{slow_cart}, the spot locations are coupled to the spot
strengths by (\ref{constraint}). One key feature of the DAE system
(\ref{slow_cart}) and (\ref{constraint}) is that it is invariant under
an orthogonal transformation. The following lemma, proved in
\ref{proof:refer_sphere}, will be used in \S\ref{sec:dynamics} for
classifying equilibria of this DAE system:

\begin{lemma}[Invariance under orthogonal transformations] 
\label{lemma:refer_sphere} Suppose that $\bm{x}_j(\sigma)$ 
for $j=1\,,\ldots,N$ is the solution to the DAE system
(\ref{slow_cart}) and (\ref{constraint}) with $\bm{x}_j(0)=
\bm{x}_j^{0}$ for $j=1,\ldots,N$. Let $\RC$ be any time-independent
orthogonal matrix. Now let $\bm{\xi}_j(\sigma)$ satisfy
(\ref{slow_cart}), (\ref{constraint}) with $\bm{\xi}_j(0)=\RC
\bm{x}_j^{0}$ for $j=1,\ldots,N$. Then, $\bm{\xi}_j(\sigma)=\RC
\bm{x}_j(\sigma)$ for all $j=1,\ldots,N$.
\end{lemma}

We emphasize that results similar to the DAE dynamics
(\ref{constraint}) and (\ref{slow_cart}) can be derived for other
RD systems. In \ref{sec:other}, we give a corresponding result
for the Schnakenberg model.

\subsection{Steady-state patterns from random initial arrangements}

To determine the dynamics and possible equilibrium spot configurations
for $N>2$ when $\E={\mathcal O}(1)$, $f$, and $\nu$ are given, we
performed numerical simulations of the DAE system (\ref{slow_cart})
and (\ref{constraint}) for both pre-specified and randomly generated
initial conditions for the spot locations. In the simulations in this
section we used $f=0.5$ and $\eps=0.02$. It is important to emphasize
that for any pattern for which the spot strengths have a common
value, it follows from (\ref{slow_cart}) and (\ref{constraint}) that
the steady-state spatial configurations of spots are independent of
$\E$, $f$, and $\nu$. In this sense, this restricted class of common
spot-strength equilibria are universal for the Brusselator, and for
other RD systems such as the Schnakenberg model. The corresponding
similar DAE dynamics for the Schnakenberg model is given in
(\ref{sslow_dyn}) of \ref{sec:other}.

To generate a set of $N$ initial points that are uniformly distributed
with respect to the surface area on the sphere, we let $h_\phi$ and
$h_\theta$ be uniformly distributed random variables in $(0,1)$ and
define spherical coordinates $\phi=2\pi h_\phi$ and
$\theta=\cos^{-1}(2h_\theta-1)$. For the initial set of $N$ points,
Newton's method was used to solve (\ref{constraint}) for the initial
spot strengths, where the initial guess for the iteration was taken
to be the two-term asymptotics (\ref{reg:s}). If the Newton iteration
scheme failed to converge, indicating that no quasi-equilibrium exists
for the initial configuration of spots, a new randomly generated
initial configuration was generated. The DAE dynamics was then
implemented by using an adaptive time-step ODE solver coupled to a
Newton iteration scheme to compute the spot strengths.

Our simulations of fifty randomly generated initial spot
configurations for the case $N=3$ suggests that a stable equilibrium
configuration consists of three equally spaced spots that lie on a
plane through the center of the sphere.  The eventual colinearity and
equal spacing between the three spot locations as time increases was
ascertained by monitoring the distances between any two spots together
with the triple product $\bm{x}_1\cdot (\bm{x}_2\times \bm{x}_3)$ at
each time step. As the slow time $\sigma$ increased, the spots became
equally spaced and the triple product tended to zero.  By using Lemma
\ref{lemma:refer_sphere}, this co-planar steady-state three-spot
configuration can be mapped by an orthogonal matrix to the standard
reference configuration of three equally spaced spots on the equator,
i.e.  $\bm{x}_j=(\cos({2\pi j/3}),\sin({2\pi j/3},0)^T$ for
$j=0,1,2$. Such a standard pattern, for which (\ref{Ge}) holds and
$\bm{S}={2\E \bm{e}/3}$, can be readily verified analytically to be a
steady-state solution for the dynamics (\ref{slow_cart}).

For $N=4$, our simulations of fifty randomly generated initial spot
configurations for the DAE dynamics suggests that the stable
equilibrium configuration generically consists of four spots centered
at the vertices of a regular tetrahedron. This was determined by
showing that as time increases, the distance between any two spots
tended to the common value $\sqrt{8/3}$ and that the volume $V_\sigma$ of
the tetrahedron formed by the spot locations, given by
\begin{equation*}
   V_\sigma = \frac{ |(\bm{x}_1-\bm{x}_4)\cdot \left[(\bm{x}_2-\bm{x}_4)\times
   (\bm{x}_3-\bm{x}_4)\right]|}{6} \,,
 \end{equation*}
tended to the volume ${8\sqrt{3}/27}$ of a regular
tetrahedron. Although our random simulations suggest that a regular
tetrahedron has a large basin of attraction for the dynamics of the
DAE system (\ref{slow_cart}) and (\ref{constraint}), it cannot
preclude the possibility of other stable steady-state configurations
with much smaller basins of attraction.

\begin{figure}[htb]
\begin{center}
\includegraphics{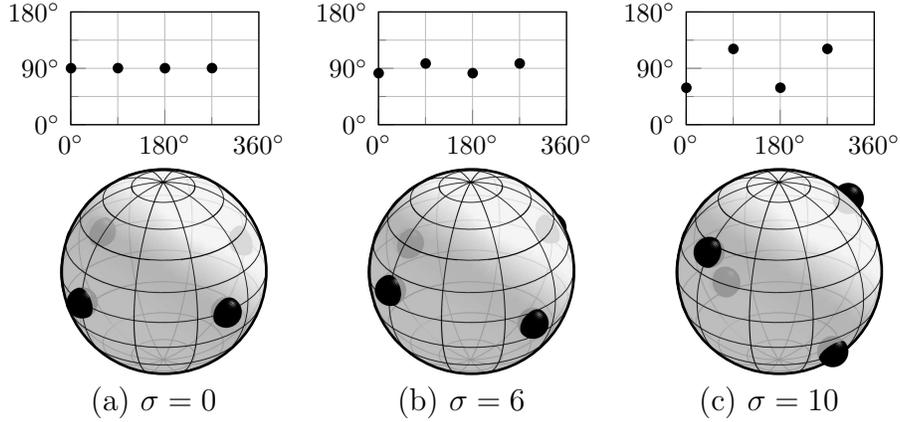}
\caption{For $f=0.5$, $\E=8$, $\eps=0.02$, four equally spaced
  spots on a ring are perturbed by a $1\%$ random perturbation in their
  locations. At $\sigma=6$ the spots have moved off of the ring,
  and at $\sigma=10$ the spots become centered at the vertices of a
  regular tetrahedron. The top subplots show the patterns in the
  $(\phi, \theta)$ plane.}
\label{ring_4}
\end{center}
\end{figure}

For any $N\geq 2$, a ring solution, consisting of $N$ equally spaced
spots on an equator of the sphere, is a steady-state solution to the
DAE system (\ref{slow_cart}) and (\ref{constraint}). For $N=3$, our
numerical computations suggest that such a ring solution is orbitally
stable to small random perturbations in the spot locations in the
sense that as time increases the perturbed spot locations will become
colinear on a nearby (tilted) ring. However, for $N\geq 4$, our
numerical simulations show that a ring solution is dynamically
unstable to small arbitrary perturbations in the spot locations on the
ring. For $N=4$, $\E=8$, $f=0.5$, and $\eps=0.02$, in
Fig.~\ref{ring_4} we show that four spots on a ring with an initial
random perturbation of $1\%$ in the spot locations will eventually
tend to a regular tetrahedron as time increases.

\begin{figure}[htb]
\begin{center}
\includegraphics[scale=1.1]{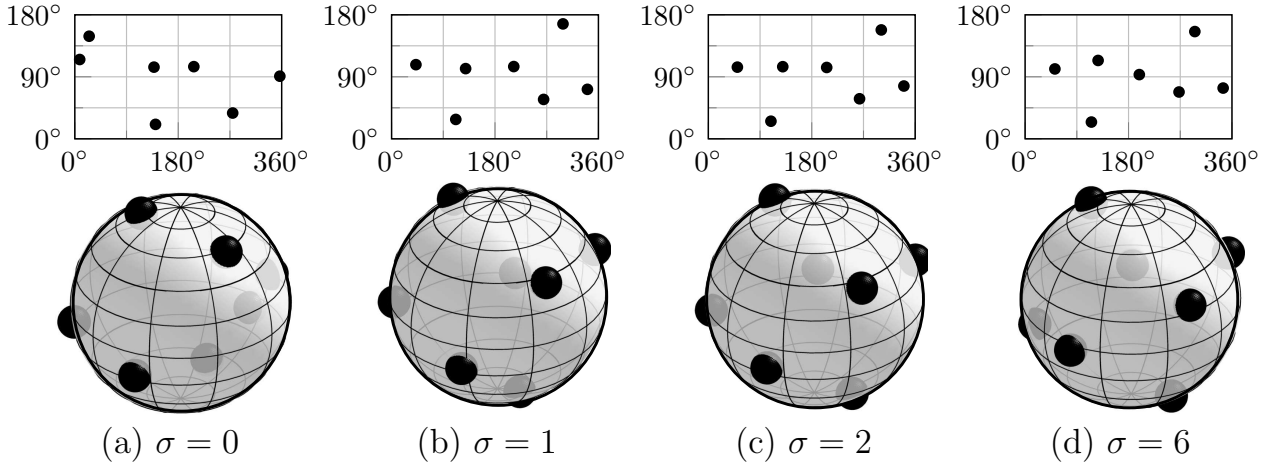}
\caption{For $f=0.5$, $\E=14$, $\eps=0.02$., seven spots, randomly
  generated, tend to an $(N-2)+2$ pattern. The pattern with
  $\sigma=6$ is near the steady-state. The top subplots
  show the patterns in the $(\phi, \theta)$ plane.}
\label{run_7}
\end{center}
\end{figure}

For $N=5$, $N=6$, and $N=7$, our numerical simulations employing fifty
randomly generated initial spot configurations for the DAE dynamics
suggests that the stable equilibrium configuration generically
consists of a pair of antipodal spots, while the remaining $N-2$ spots
are equally spaced on the mid-plane between these two spots. We refer to
such patterns as $(N-2)+2$ patterns. The diagnostics used to form this
conclusion are as follows. For each initial condition, we solved the
DAE dynamics until a steady-state was reached. From this steady-state
configuration two antipodal spots, labelled by $\bm{x}_1$ and
$\bm{x}_2=-\bm{x}_1$, were identified from a dot product.  We
arbitrarily chose $\bm{x}_1$ to map to $\bm{\xi}_1=(0,0,1)^T$. We then
chose any one of the other $N-2$ spots locations, such as $\bm{x}_3$,
and map $\bm{x}_3$ to $\bm{\xi}_3=(1,0,0)^T$. We define $\RC$ to be
the orthogonal matrix where the first row is $\bm{x}_3$, the second
row is ${(\bm{x}_1\times \bm{x}_3)/|\bm{x}_1 \times \bm{x}_3|}$, and
then third row is $\bm{x}_1$. With this choice for the matrix $\RC$,
we found that the computed steady-state points $\bm{x}_j$, for
$j=1\ldots,N$, can be mapped to the standard reference configuration
for an $(N-2)+2$ pattern consisting of spots at $(0,0,1)$ and
$(0,0,-1)$, and $N-2$ spots equally spaced on the equator
$\theta={\pi/2}$ with one of these spots at $(1,0,0)^T$. This 
mapping technique was fully automated and allowed us to identify the
final steady-state pattern computed from the DAE dynamics. For $N=7$,
the numerical results shown in Fig.~\ref{run_7} illustrate the
formation of the $(N-2)+2$ pattern from a random initial condition for
the parameter set $f=0.5$, $\E=14$, and $\eps=0.02$. The $(N-2)+2$
structure is evident from Fig.~\ref{run_7}(d), which is close to the 
steady-state pattern. When $N=6$, the $(N-2)+2$ pattern is simply
an octahedron.

For an $(N-2)+2$ pattern, the two anitipodal spots have strength $S_p$
while the remaining $(N-2)$  equally-spaced spots on the equator have 
strength $S_c$. By partitioning the
Green's matrix in (\ref{constraint}) into a cyclic $(N-2)\times (N-2)$
sub-block consisting of spot interactions on the ring, we can derive
after some algebra from (\ref{constraint}) that $S_c$ satisfies the
scalar nonlinear algebraic equation
\begin{equation} \label{N-2+2} 
\begin{gathered}
  S_c - \frac{2\nu}{N} S_c \left[\log(N-2) - \frac{(N-2)}{2}\log{2} \right]
  + \frac{2\nu}{N} \left[ \chi(S_c)-\chi(S_p)\right]-\frac{2\E}{N}=0\,, \\
  S_p=\E- \frac{(N-2)}{2}S_c\,.
  \end{gathered}
\end{equation}
For all of our numerical DAE computations for $N=5,6,7$, we verified that
the spot strengths for the steady-state pattern satisfied (\ref{N-2+2}).

\begin{figure}[htb] \centering
\includegraphics[scale=1.1]{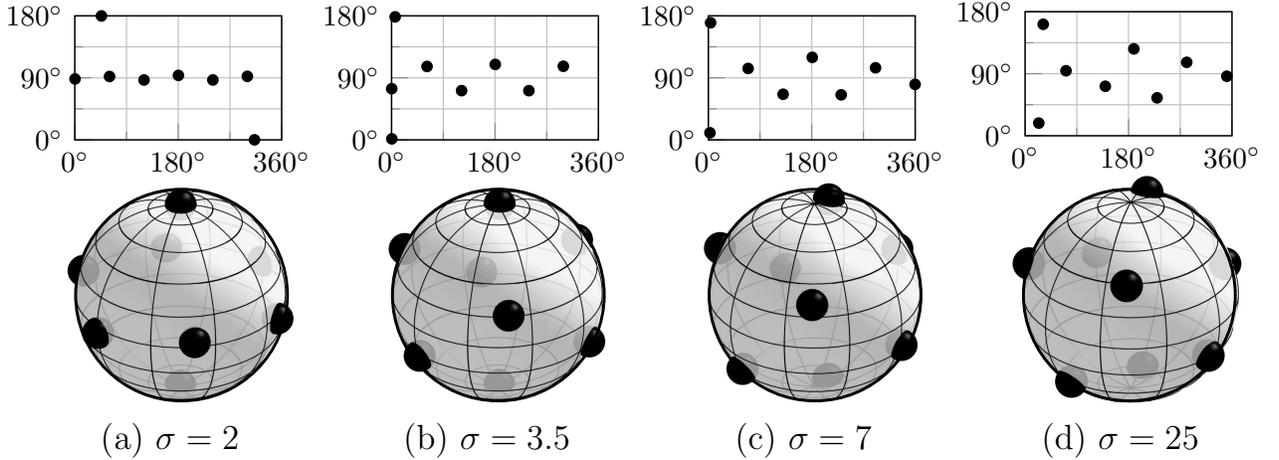}
\caption{For $f=0.5$, $\E=16$, and $\eps=0.02$. Eight spots in a
  standard $(N-2)+2$ pattern undergo a $1\%$ random perturbation at
  time $\sigma=0$. The initial $(N-2)+2$ pattern is found to be
  unstable. The pattern for $\sigma=25$ is near the steady-state
  pattern. The top subplots display the patterns in the $(\phi,
  \theta)$ plane. \label{run_8}}
\end{figure}

Our numerical results show that the $(N-2)+2$ pattern for $N\geq 8$ is
unstable. This is illustrated for $N=8$ in Fig.~\ref{run_8} where we
took an initial $1\%$ random perturbation in the spot locations.
However, unlike the cases for $N < 8$ where the $(N-2)+2$ patterns
were visually discernible, the final steady-state pattern in
Fig.~\ref{run_8}(d) is no longer clear. For a general steady-state
configuration of $N$ points, we now propose an algorithm to
rotate the sphere so that the symmetries are apparent.

\begin{figure}[htb] \centering
\includegraphics[scale=1.2]{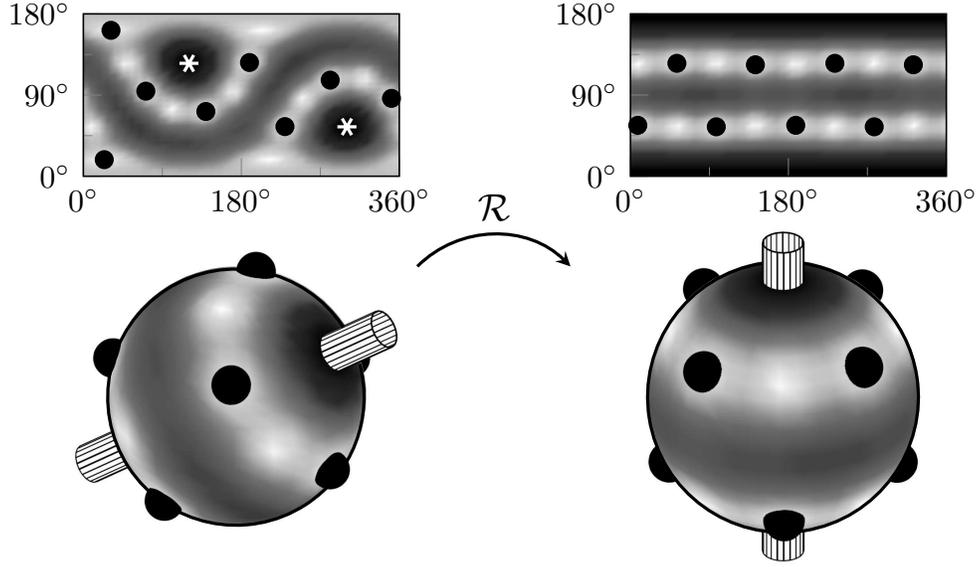}
\caption{(Left) The same as Fig.~\ref{run_8}(d), but the shading on
  the sphere and top $(\phi, \theta)$ plane show values of
  $\mathscr{D}(\bm{x})$, with light and dark shading for small and
  large values, respectively. The two asterisks (global maxima)
  indicate a better location to place the polar axis of the sphere
  (marked by a cylinder). (Right) After an orthogonal transformation,
  $\mathcal{R}$, the rotated sphere in the (new) $(\bar{\phi},
  \bar{\theta})$-plane shows two rings of four
  spots. \label{fig:rotate}}
\end{figure}

Let $\Delta(\bm{x}, \bm{y}) > 0$ be the great circle distance along
the geodesic connecting the two points, $\bm{x}$ and $\bm{y}$, on the
sphere. To each point, $\bm{x}$, on the sphere, we compute
\begin{equation}
  \mathscr{D}(\bm{x}) = \sum_{i=1}^N \Bigl[ \Delta(\bm{x},
    \bm{x}_i)^\alpha + \Delta(\textrm{antipodal of } \bm{x},
    \bm{x}_i)^\alpha\Bigr].
\end{equation}
That is, $\mathscr{D}(\bm{x})$ is a measure of the closeness of
$\bm{x}$ and its antipodal point to the set of spots. The value of
$\alpha > 0$ is a weighting parameter designed to penalize distance to
the spots (we choose $\alpha = 0.5$). Let $\bm{x}^*$ be an extremum
(either local or global) of $\mathscr{D}$ on the sphere. We observe
that by rotating the sphere so that the new north and south poles are
oriented along $\bm{x}^*$ and its antipodal point, the symmetry
patterns often become clear in the new $(\bar{\theta}, \bar{\phi})$
plane. This is shown in Fig.~\ref{fig:rotate} for the spot pattern in
Fig.~\ref{run_8}(d), which is now recognized as forming what we refer
to as a $45^{\circ}$ ``twisted cuboid'': two parallel rings containing
four equally-spaced spots, with the rings symmetrically placed above
and below the equator, and with the spots phase shifted by $\bar{\phi}
= 45^\circ$ between each ring. However, since the distance between the
two parallel planes is not the same as the minimum distance between
any two neighbouring spots on the same ring, the untwisted shape does
not form a true cube. Our computations yield that the perpendicular
distance between the two planes is $\approx 1.12924$ as compared to a
minimum distance of $\approx 1.1672$ between neighboring spots on the same
ring. The ratio of this minimum to perpendicular distance is
approximately $0.967$. This yields that the rings are at latitudes
$\theta\approx 55.6^{\circ}$ and $\theta\approx 124.4^{\circ}$ (see the subplot
in Fig.~\ref{fig:rotate}).

Further numerical simulations of randomly generated eight-spot
patterns suggests that the stable equilibrium pattern is generically
the $45^{\circ}$ degree twisted cuboid described above. Our numerical
results also show that an untwisted cuboid is unstable to small random
perturbations, and that a cuboid with initial twist angle $\omega$
will tend to a $45^{\circ}$ twisted cuboid as time increases. 

\subsection{A ring pattern with a polar spot: prediction of a triggered instability}

Next, for $N\geq 3$ we consider an initial pattern with $(N-1)$ spots
equidistantly spaced on a ring of constant latitude $\theta(0)$
together with a polar spot centered at $\theta=0$. For this special
$(N-1)+1$ pattern, we can reduce the DAE system
(\ref{constraint}) and (\ref{slow_cart}) to a scalar ODE for the
latitude of the ring coupled to a single nonlinear algebraic equation
for the common spot strength for the spots on the ring. For this type
of pattern we will predict the occurrence of a dynamically triggered
spot-splitting instability.

In terms of spherical coordinates, we have for the $N-1$ spots on the ring
at time $\sigma=0$ that $\theta_j(0)=\theta(0)$ and $\phi_j(0)=
{2\pi (j-1)/(N-1)}$ for $j=1,\ldots,N-1$. For the polar spot, we have
$\theta_N(0)=0$. From (\ref{slow_dyn_e}), it is readily shown that for
all time, $\sigma\geq 0$
\begin{equation*}
    \phi_{j}(\sigma)=\phi_{j}(0)\,; \qquad \theta_{j}(\sigma)=\theta_{c}(\sigma)
\,, \quad j=1,\ldots,N-1 \,; \qquad \theta_{N}(\sigma)=0 \,,
\end{equation*}
where $\theta_{c}(\sigma)$, with $\theta_{c}(0)=\theta(0)$, is the common
latitude of the $N-1$ spots on the ring. For this pattern, the spot 
spot-strengths are $\bm{S}=(S_c,\ldots,S_c,S_N)^T$, where
$(N-1)S_c + S_N = 2\E$. 

The dynamics of the $(N-1)+1$ spot pattern is characterized in terms
of an ODE for $\theta_{c}(\sigma)$ coupled to a nonlinear algebraic
equation for $S_c=S_c(\theta_c)$. By partitioning the Green's matrix
in (\ref{constraint}) into a cyclic $(N-1)\times (N-1)$ sub-block consisting
of spot interactions on the ring, we readily obtain from
(\ref{constraint}) that $S_c$ satisfies the scalar nonlinear
algebraic equation 
\begin{subequations}  \label{cf:N-1+1}
\begin{equation}
  \T(S_c) \equiv NS_c + \nu \left[\chi(S_c)-\chi(S_N) + S_c
   \left( 2(N-1)L -\kappa_N\right)\right] - 2\E \left(1+\nu L\right)=0 \,,
\end{equation}
where $S_N=2\E-(N-1)S_c$.  Here $L=L(\theta_c)$ is the common
value $L=\log|{\bm x}_j-{\bm x}_N|$ for $j=1,\ldots,N-1$, and
$\kappa_N$ is the eigenvalue of the $(N-1)\times (N-1)$
cyclic sub-block of ${\mathcal G}$ with corresponding $N-1$
dimensional eigenvector $(1,\ldots,1)^T$. A calculation yields
that
\begin{equation}
    L= \log\left[ 2 \sin\left({\theta_c/2}\right)\right] \,, \quad
    \kappa_N = \sum_{\substack{j=1 \\ j \neq k}}^{N-1} \log|\bm{x}_j-\bm{x}_k|
   = \log(N-1) + (N-2)\log\left(\sin\theta_c\right)\,.  \label{ln_kn}
\end{equation}
\end{subequations}

To determine the ODE for $\theta_c$, we set $\theta_{j}=\theta_c$ for
$j=1,\ldots,N-1$ in (\ref{slow_dyn_e}) to obtain that
\begin{equation}
\dd{\theta_c}{\sigma} = -(N-2) \frac{ S_c}{\mathcal{A}(S_c)} 
 \cot\theta_c - \frac{S_N}{{\mathcal A}(S_N)} \cot\left({\theta_c/2}\right)\,,
  \qquad \theta_c(0)=\theta(0) \,, \label{N-1+1:ode}
\end{equation}
where $S_N=2\E-(N-1)S_c$. The DAE system for this pattern is
to solve (\ref{N-1+1:ode}) together with the constraint $\T(S_c)=0$ of
(\ref{cf:N-1+1}), which yields $S_c=S_c(\theta_c)$. As a remark, if we
set $N=2$ in (\ref{cf:N-1+1}) and (\ref{N-1+1:ode}) we obtain
$S_c=S_N=E$, and readily recover the two-spot dynamics of Lemma
\ref{lemma:two-spot}.

Since ${\mathcal A}(S_c)<0$ and ${\mathcal A}(S_N)<0$, we observe from
(\ref{N-1+1:ode}) that $\theta_c^{\prime}>0$ for
$0<\theta_c<{\pi/2}$, but $\theta_{c}^{\prime}<0$ as $\theta_c\to
\pi^{-}$.  As such, (\ref{N-1+1:ode}) will have a steady-state at some
$\theta_{ce}$ satisfying ${\pi/2}<\theta_{ce}<\pi$.

\begin{figure}[htb]
\begin{center}
  \includegraphics{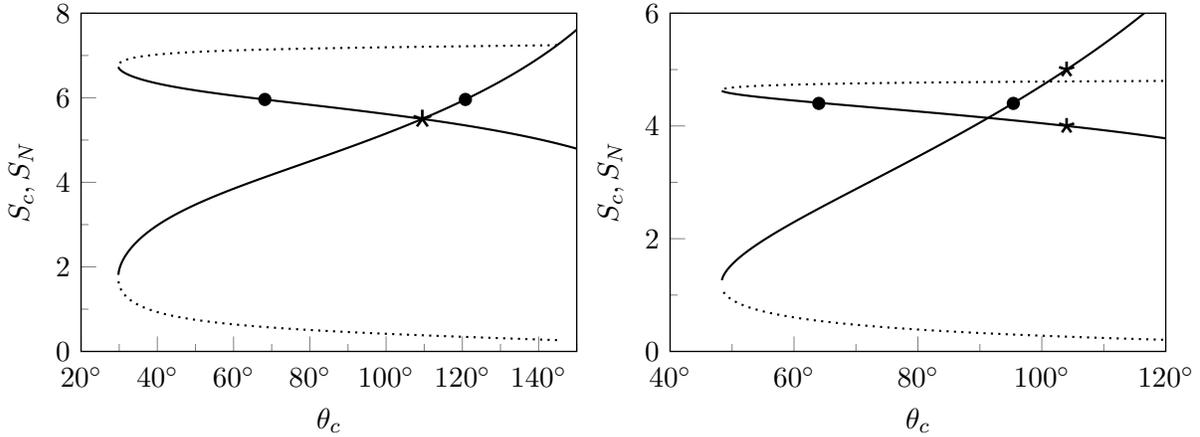}
\caption{Plot of the common spot strength $S_c$ for the $N-1$ spots
  on a ring (tight C-shaped curve) and the spot strength $S_N$ for
  the polar spot (open C-shaped curve) versus the ring latitude
  $\theta_c$ (in degrees), as computed from (\ref{cf:N-1+1}). The
  upper (lower) branch of the $S_c$ curve corresponds to the lower
  (upper) branch of the $S_N$ curve. The dashed portions of these
  curves represent quasi-equilibria that are unstable on an ${\mathcal
    O}(1)$ time-scale since $S_N={\mathcal O}(\nu)$ (see \S
  \ref{sec:qe_instab}). The unique steady-state of the slow dynamics
  (\ref{N-1+1:ode}) is indicated by ${\star}$. Left panel: $f=0.5$,
  $\eps=0.02$, $\E=11$, and $N=4$.  Right panel: $f=0.6$, $\eps=0.02$,
  $\E=14.5$ and $N=7$. In these panels, the spot self-replication
  thresholds $\Sigma_2(0.5)\approx 5.96$ and $\Sigma_2(0.6)\approx
  4.41$ are indicated by ${\bullet}$. From the right panel, for
  $\theta_{c}(0)=80^{\circ}$, we predict that the polar
  spot with spot strength $S_N$ will undergo a dynamically triggered
  spot self-replication instability before reaching the steady-state.}
\label{fig:npolar}
\end{center}
\end{figure}

In the left and right panels of Fig.~\ref{fig:npolar} we plot the
solutions $S_c$ and $S_N$ to (\ref{cf:N-1+1}) as a
function of $\theta_c$ for two different parameter sets. We observe
that there is a minimum latitude, depending on $\E$, $N$, and $f$, for
which quasi-equilibria can exist, which yields a saddle-node
bifurcation structure. In these figures, the
upper (lower) branch of the $S_c$ curve corresponds to the lower
(upper) branch of the $S_N$ curve. The dashed portions of these curves
are quasi-equilibria that are unstable on an ${\mathcal O}(1)$
time-scale since $S_N={\mathcal O}(\nu)$ (see
\S\ref{sec:qe_instab}). In these figures the unique steady-state,
$\theta_{ce}$, of the slow dynamics (\ref{N-1+1:ode}) is indicated by
a star $({\star})$, while the spot self-replication threshold is
marked by a circle $(\bullet)$.

The implication of these results for spot dynamics is as follows.  For
any initial value $\theta_c(0)<\theta_{ce}$, (\ref{N-1+1:ode}) yields
$\theta_{c}^{\prime}(\sigma)>0$, so that $\theta_c(\sigma)$ increases
monotonically towards $\theta_{ce}$. In this case, $S_c$ decreases
while $S_N$ increases along the solid curves in Fig.~\ref{fig:npolar}
until the steady-state is reached. Alternatively, if
$\theta_c(0)>\theta_{ce}$, then $S_c$ increases and $S_N$ decreases
along the solid curves in Fig.~\ref{fig:npolar} until reaching the
steady-state. If at $\sigma =0$ or at any $\sigma>0$ either $S_c$ or $S_N$
exceeds the threshold $\Sigma_2(f)$, we predict that a spot self-replication 
event will occur. If the threshold is exceeded only at a later time $\sigma>0$,
we refer to this instability as a {\em dynamically triggered instability}.

The plots in Fig.~\ref{fig:npolar} reveal several possible dynamical
behaviors. First, consider the parameter set $f=0.5$, $N=4$, $\E=11$,
and $\eps=0.02$, corresponding to the left panel of
Fig.~\ref{fig:npolar}. For an initial angle satisfying
$68^{\circ}<\theta_c(0)<121^{\circ}$, we observe that no
spot-splitting can occur and $\theta_{c}\to\theta_{ce}\approx
109.3^{\circ}$ as $\sigma\to \infty$. For $\theta_c(0)<68^{\circ}$,
but above the saddle-node value, we have $S_c>\Sigma_2(0.5)$ and so
predict that the $3$ spots on the ring will undergo a spot
self-replication process beginning at $\sigma=0$. Alternatively, for
$\theta_{c}(0)>121^{\circ}$, we predict that the polar spot will
undergo splitting starting at $\sigma=0$. For the parameter set
$f=0.6$, $N=7$, $\E = 14.5$, and $\eps=0.02$, corresponding to the
right panel of Fig.~\ref{fig:npolar}, we observe that a dynamically
triggered instability can occur for the polar spot. To illustrate
this, suppose that $\theta_c(0)=80^{\circ}$. Then, from the right
panel of Fig.~\ref{fig:npolar}, it follows that $S_N$ will exceed the
spot-splitting threshold $\Sigma_2(0.6)\approx 4.41$ before reaching
the steady-state value. Thus, we predict that the slow dynamics will
trigger, at some later time, a spot self-replication event
for the polar spot.

\section{Discussion}\label{sec:disc}

Asymptotic analysis has been used to derive a DAE system
(\ref{constraint}) and (\ref{slow_cart}) characterizing the slow
dynamics of localized spot solutions for the Brusselator on the sphere. When the quasi-equilibrium spot
solution is linearly stable to ${\mathcal O}(1)$ time-scale
instabilities, the system describes the
motions of a collection of $N$ spots on a long time-interval of order
${\mathcal O}(\eps^{-2})$. Numerical simulations of the DAE system
with random initial spot locations has identified stable spatial
configurations with large basins of attraction for equilibrium spots
with $2\leq N\leq 8$. For the case $N=8$, such a stable spot pattern is
a $45^{\circ}$ twisted cuboid, consisting of four equally spaced spots
on two parallel rings, with spots on the two rings phase-shifted by
$45^{\circ}$, and where the rings are at the approximate latitudes
$55.6^{\circ}$ and $124.4^{\circ}$.

Although our results do not address the fundamental question of how
many localized spots will form starting from a small random
perturbation of the spatially uniform state, our stability results in
\S \ref{sec:qe_instab} can be used to give leading-order-in-$\nu$
bounds on the minimum and maximum number of spots in a stable
steady-state pattern. To leading-order in $\nu$, we showed in \S
\ref{sec:qe_instab} that stable spot patterns are those for which all
individual spot strengths, $S_j$, tend to the common value $S_c$ as
$\nu\to 0$ [see \eqref{reg:s}].  Using this leading order estimate,
the $N$-spot pattern is stable to spot self-replication when $N$ is
large enough so that $S_c<\Sigma_2(f)$, Moreover, it is stable to a
competition or overcrowding instability when $N$ is small enough so
that $S_c>\sqrt{\nu d_0}$ [see \S\ref{sec:type3}]. This yields the
following bounds in the limit $\nu\to 0$ on the number $N$ of stable
steady-state spots:
\begin{equation}
     \frac{2\E}{\Sigma_{2}(f)} < N < \frac{2\E}{\sqrt{\nu}} 
   \frac{f}{\sqrt{b(1-f)}} \,. \label{bound}
\end{equation}
For the parameter set $\eps=0.075$, $f=0.8$, and $\E=4.0$ of
Fig.~\ref{fig:evo}, we use $\Sigma_2(0.8)\approx 2.28$ to calculate
$3.51<N<10.36$ from (\ref{bound}). The computed pattern in
Fig.~\ref{fig:evo} had 6 spots. We remark that the bounds in
(\ref{bound}) will be tighter, and hence more useful, for smaller
values of $f$.

DAE systems for slow spot dynamics, similar to (\ref{constraint}) and
(\ref{slow_cart}) for the Brusselator, can also be derived for other
RD systems. For example, in \ref{sec:other}, we present analogous
results for the Schnakenberg model. The primary feature that is needed
to apply the analysis herein is that the outer approximation for
the quasi-static inhibitor concentration $v$ (i.e. the long range solution 
component) must satisfy a linear elliptic problem
on the sphere of the form $\dell_S v-\kappa v = A+\sum_{j=1}^{N} S_j
\delta(\bm{x}-\bm{x}_j)$, for some $\kappa\geq 0$ and constant $A$.

Finally, we compare our result for spot dynamics with the
well-known results for the dynamics of a collection of point vortices
centered at $\bm{x}_i$, for $i=1,\ldots,N$, on the sphere for Euler's 
equations. For $N$ such point vortices of strength 
$\Gamma_i$, for $i=1,\ldots,N$, the ODE point vortex dynamics are
(cf.~\cite{Bogomolov}, \cite{Newton2001})
\begin{equation}
   \bm{x}_j^{\prime} = \frac{1}{2\pi} \sum_{\substack{i=1 \\ i \neq j}}^N  
\Gamma_i \frac{\bm{x}_i \times \bm{x}_j}{|\bm{x}_i-\bm{x}_j|^2}\,, \qquad
  j=1,\ldots,N \,, \label{p:vort_1}
\end{equation}
subject to $\sum_{i=1}^{N}\Gamma_i=0$. In terms of spherical
coordinates, (\ref{p:vort_1}) for $j=1,\ldots,N$ becomes
\begin{subequations} \label{p:vort_2}
\begin{align}
\dd{\theta_j}{t} &= -\frac{1}{4\pi} \sum_{\substack{i=1 \\ i \neq j}}^N 
\frac{\Gamma_i}{1-\cos\gamma_{ij}} \sin\theta_i \sin(\phi_j-\phi_i) \,,\\
\sin\theta_j \dd{\phi_j}{t} &= \frac{1}{4\pi} 
\sum_{\substack{i=1 \\ i \neq j}}^N \frac{\Gamma_i}{1-\cos\gamma_{ij}}
  \left[\sin\theta_j\cos\theta_i-\cos\theta_j\sin\theta_i\cos(\phi_j-\phi_i)
 \right] \,, 
\end{align}
\end{subequations}
where $\gamma_{ij}$ is the angle between $\bm{x}_i$ and $\bm{x}_j$. In
contrast to our result for slow spot dynamics, the ODE system
(\ref{p:vort_2}) is Hamiltonian. This structure has been used for
analyzing (\ref{p:vort_2}) for specific problems such as, the
stability of a latitudinal ring of vortices (cf.~\cite{Boatto}), the
integrable $3$-vortex problem (cf.~\cite{Newton1998}), and
characterizing relative equilibria of point vortex configurations
(cf.~\cite{Newton2011}).

Our asymptotic result (\ref{slow_dyn_e}) and (\ref{slow_cart}) for
slow spot dynamics differs in at least two key aspects from the point
vortex dynamics of (\ref{p:vort_1}) and (\ref{p:vort_2}). Firstly, in
(\ref{slow_dyn_e}) and (\ref{slow_cart}), the spot strengths $S_j$ are
not pre-specified, but instead are coupled to the slow dynamics by the
nonlinear algebraic constraint (\ref{constraint}). This leads to an
ODE-DAE system for slow spot dynamics. In contrast, for the point
vortex problem, the vortex strengths $\Gamma_i$ are arbitrary, subject
only to the constraint that $\sum_{i=1}^{N}\Gamma_i=0$. Secondly, the
results in (\ref{slow_dyn_e}) and (\ref{slow_cart}) are asymptotically
valid only when the quasi-equilibrium profile in (\ref{quasi_eq}) is
linearly stable to ${\mathcal O}(1)$ time-scale instabilities. One
such instability leads to the triggering of a nonlinear spot
self-replication event, and this instability occurs whenever the local
spot strength $S_j$ exceeds a threshold $\Sigma_2=\Sigma_2(f)$
(cf.~\cite{rozada2014}). A discussion of these instabilities and their
implications on slow spot dynamics was discussed in \S
\ref{sec:qe_instab}. There is no comparable phenomena for the point
vortex problem.

\subsection{Open problems} 

We now discuss several possible directions that warrant further
investigation. 

\subsubsection{Equilibria and the Green's matrix}
One central issue concerns the Green's matrix, $\G$,
appearing in the nonlinear algebraic system \eqref{constraint}. When
the spots are distributed in such a way that $\bm{e}$ is an
eigenvector of $\G$, we have been able to expose the bifurcation
structure of the solutions for the spot strengths (see
\S\ref{sec:qe_overview}). For this case, there is a solution to
(\ref{constraint}) where the spots have a common spot strength, and
the number of distinct bifurcation points (in $\E$) from this
symmetric solution branch in the $\E={\mathcal O}(\nu^{1/2})$ regime
is the number of distinct eigenvalues of $\G$ in the subspace
orthogonal to $\bm{e}$. Although it is easy to verify that $\bm{e}$ is
an eigenvalue of $\G$ for some simple spatial arrangements of spot
patterns such as, equally-spaced spots on a ring of constant latitude,
spots centered at the vertices of any platonic solid (see Table 1 of
\cite{rozada2014}), or eight spots forming a twisted cuboid, it is an
open problem to numerically classify all spot configurations for which
$\bm{e}$ is an eigenvector of $\G$. For larger values of $N$, it was
shown in Table 2 of \cite{rozada2014} that the elliptic Fekete points,
defined as the point set that {\em globally} minimizes the discrete
logarithmic energy $V\equiv -\sum\sum_{i\neq j}
\log|\bm{x}_i-\bm{x}_j|$ with $|\bm{x}_i|=1$, generates a Green's
matrix $\G$ for which $\bm{e}$, as measured in the $L_2$ norm, is
rather close to an eigenvalue of $\G$. We remark that if we set
$S_j=S_c$ for $j=1,\ldots,N$ in (\ref{slow_cart}), then any stable
steady-state solution of (\ref{slow_cart}) must correspond to a local
minimum of the discrete logarithmic energy. By calculating the
discrete logarithmic energy of our $45^{\circ}$ twisted cuboid,  and
then examining Table 1 of \cite{SSP}, we have verified that our 8-spot
twisted cuboid is indeed an elliptic Fekete point set and not just a
local minimum of the discrete logarithmic energy. These 
observations suggest that it would be interesting to carefully examine
the relation between elliptic Fekete points and equilibria of
(\ref{constraint}) and (\ref{slow_cart}).  

We further remark that when $\bm{e}$ is an eigenvector of $\G$, the
steady-state spot locations for an $N$-spot pattern, having spots of a
common spot strength, are independent of the parameters in the RD
model. A similar universality result holds for common spot strength
patterns in the Schnakenberg model (see (\ref{thm:sdyn}) of
\ref{sec:other}).

Another open problem is to use numerical bifurcation software to
path-follow the small amplitude weakly nonlinear spatial patterns,
which emerge from a Turing bifurcation when $\epsilon=O(1)$,
  into the regime $\epsilon\ll 1$ of localized spot patterns studied
  in this paper. In particular, as $\epsilon$ is varied, do our
  localized spot patterns arise from subcritical bifurcations of the
  weakly nonlinear amplitude equations?

\subsubsection{Bifurcations and imperfection-sensitivity}

However, when $\bm{e}$ is not an eigenvalue of $\G$, our numerical
investigation for $N=3$ of the solution set to the constraint
(\ref{constraint}), has shown the qualitatively new result that the
leading-order-in-$\nu$ bifurcation diagram in the $\E={\mathcal
  O}(\nu^{1/2})$ regime is imperfection sensitive to small
perturbations resulting from higher order in $\nu$ terms.  This
imperfection sensitivity of the bifurcation structure of (\ref{constraint})
when $\bm{e}$ is not an eigenvalue of $\G$ is a qualitatively
new result in the construction of spot-type patterns. Previous
asymptotic constructions of asymmetric spot-type patterns for other RD
models such as the Gierer-Meinhardt, Gray-Scott, or
Schnakenberg models in planar 2-D domains (see \cite{survey_Wei:2008}
for a survey), were based on a leading-order-in-$\nu$ theory, and
hence the effect of higher order in $\nu$ terms were not considered.
For $\nu$ small and any $N>2$, it would be interesting to provide an
asymptotic analysis of imperfection sensitivity for these other RD models.

An intriguing question concerns identifying and then classifying the
steady-state spot configurations of the DAE system (\ref{constraint})
and (\ref{slow_cart}), as was studied in
\S\ref{sec:dynamics}. Although the patterns for $N \leq 8$ were
relatively easy to recognize, it would be interesting to devise a
numerical algorithm based on ideas from group theory to classify into
symmetry groups any stable steady-state spot patterns on the sphere
when $N > 8$. We note furthermore that since the DAE system does not
appear to be a gradient flow, it would also be interesting to explore
whether it can admit irregular dynamics for some special initial
conditions, or for larger values of $N$ than we have examined. An
additional open problem is to analytically perform a stability
analysis of steady-state solutions of the DAE system
(\ref{constraint}) and (\ref{slow_cart}).


\subsubsection{Comparisons with full numerical simulations} 
In order to benchmark the range of validity in $\epsilon$ of the
asymptotic slow-spot dynamics, we would require full numerical
simulations of the Brusselator model (1) over long time
intervals. Indeed, for the simpler case of the Gray-Scott model posed
on a rectangular domain, results from a related DAE system were
favorably compared in [6] with full numerical results computed by a
finite-element software package. However, the analogous study for the
sphere and for general curved surfaces remains an open question.

Currently, our method for computing the patterns shown in
Fig.\ref{fig:evo} relies on an explicit time-stepping scheme using the
closest-point method (\emph{c.f.} references in
\cite{rozada2014}). However, such explicit schemes are inadequate for
obtaining the accuracy and time-scales necessary to validate the
$\epsilon \to 0$ limit, and one would require the development of an
implicit numerical solver. For example, one possible numerical
approach would be to use a spectral method, tailored for the sphere,
coupled to implicit-explicit (IMEX) scheme for the time-stepping. The
development of such a code, which could be used for comparisons with
the DAE system, is beyond the scope of this paper, but we highlight
this task as an important problem for future work.

\section*{Acknowledgements}
PHT thanks Lincoln College, Oxford and the Zilkha Trust for generous
funding. MJW gratefully acknowledges grant support from NSERC. We are
grateful to Prof.~Paul Matthews of Nottingham University regarding possible
stable spot patterns for $N=8$ spots, and to Prof.~Stefanella
Boatto for discussions about the point-vortex problem.

\begin{appendix}
\newcommand{\newsection}[1]{{\setcounter{equation}{0}}\section{#1}}
\renewcommand{\theequation}{\Alph{section}.\arabic{equation}}
\section{Non-dimensionalization of the Brusselator} \label{sec:brus}

The standard form for the Brusselator RD model is (cf.~\cite{Prigogine1968})
\begin{equation}
\partial_T U = \eps_0^2 \dell_S U + \hat{E} - (B + 1) U + U^2 V \,, \quad
\partial_T V = D \dell_S V + BU - U^2 V \,, \label{b_sys}
\end{equation}
where $\eps_0^2\equiv {D_U/L^2}$, $D\equiv {D_V/L^2}$, and $L$ is the
radius of the sphere. Here $\dell_S$ is the surface Laplacian for the
unit sphere. We consider the singularly perturbed limit $\eps_0\to 0$
for which $D={D_v/L^2}={\mathcal O}(1)$ as $\eps_0\to 0$. In
\cite{rozada2014} it was shown that localized spot patterns for
(\ref{b_sys}), characterized by localized regions where $U={\mathcal
  O}(\eps_0^{-1})$, exist when $\hat{E}={\mathcal O}(\eps_0)$. We
scale (\ref{b_sys}) so that the amplitude of the spots is ${\mathcal
  O}(1)$ as $\eps_0\to 0$. In terms of the new variables $t$, $u$, and
$v$, defined by
\begin{equation*}
    T=\frac{t}{B+1} \,, \qquad U = \frac{\sqrt{(B+1)D}}{\eps_0} u \,, \qquad
  V = \frac{B}{\sqrt{(B+1)D}} \eps_0 v \,,
\end{equation*}
we get that (\ref{b_sys}) reduces to (\ref{full_all}), where $f$,
$\tau$, $\eps$, and $\E={\mathcal O}(1)$ in (\ref{full_all}) are
defined by
\begin{equation}
   \eps \equiv \frac{\eps_0}{\sqrt{B+1}} \,, \qquad \tau
   \equiv\frac{(B+1)}{D} \,, \qquad f \equiv \frac{B}{B+1} \,, \qquad E
   \equiv \frac{\hat{E}}{\sqrt{(B+1)D}\eps_0} \,. \label{b_fin:2}
\end{equation}
Our non-dimensionalization of the Brusselator so that $v$ has
unit diffusivity is slightly different than that used in
\cite{rozada2014}. However, the system studied in \cite{rozada2014}
can be readily mapped to (\ref{full_all}).

\section{Further details of the leading-order inner solution} \label{appendix:core}

\begin{figure}[tb] \centering
\includegraphics[width = 0.48\textwidth,height=4.8cm]{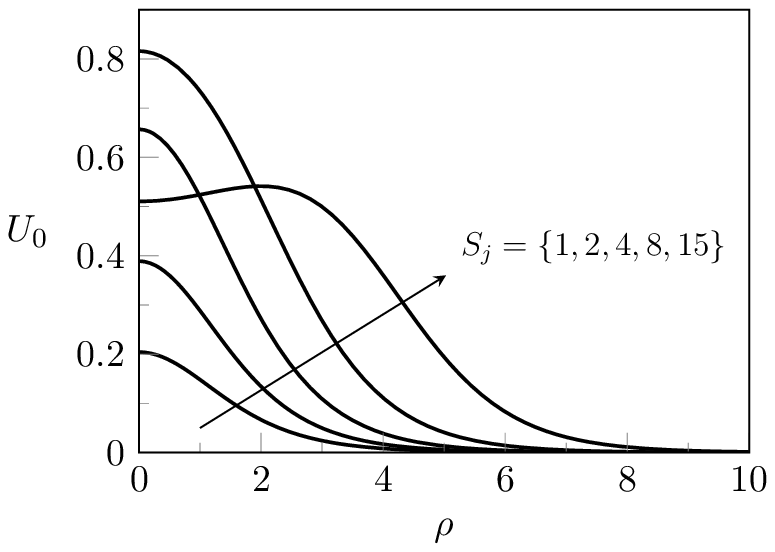}
\caption{$U_0=U_{j0}(\rho)$ for $f = 0.3$ and $S_j = \{1, 2, 4, 8,
  15\}$. As $S_j$ increases, $U_{j0}$ develops a volcano
  profile. \label{UV}}
\end{figure}

\begin{figure}[htb]
\begin{center}
\includegraphics{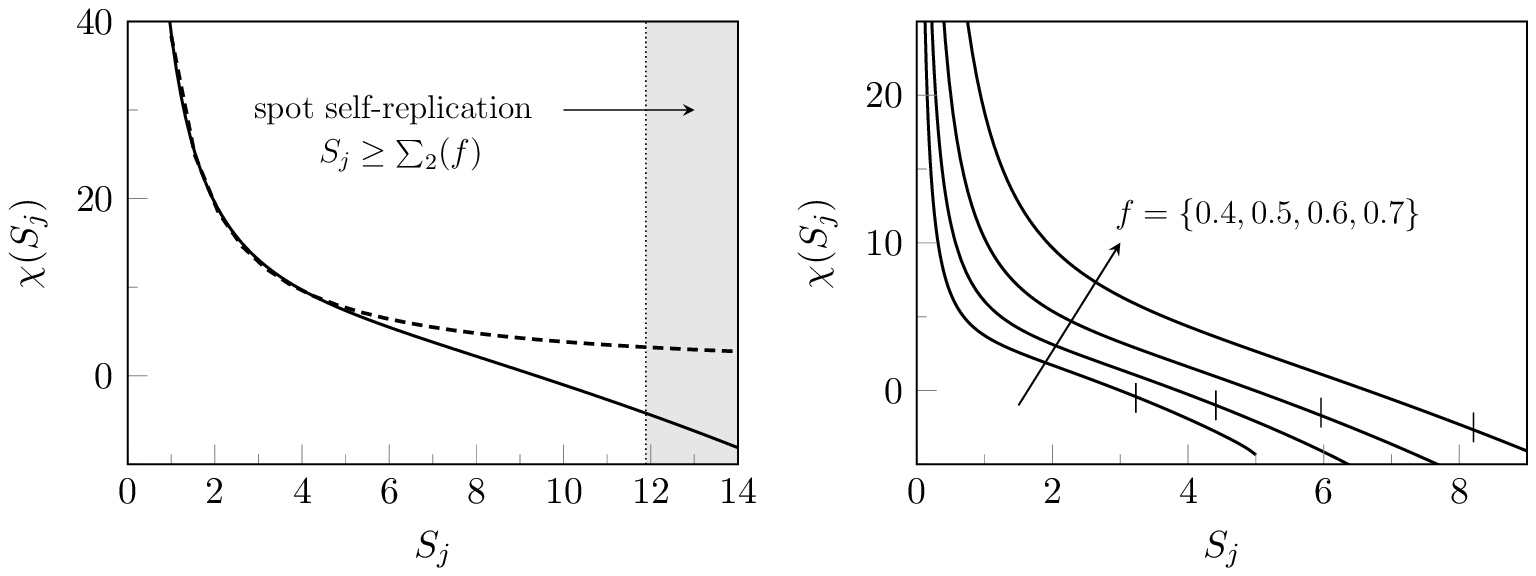}
\caption{Left: $\chi$ versus $S_j$ for $f=0.3$ (heavy solid curve).
  The dashed curve is the asymptotic result $\chi \sim b(1 -
  f)/(S_jf^2)$ as $S_j\to 0$ with $b\approx 4.934$.  Right: $\chi$
  versus $S_j$ for $f=0.4$, $f=0.5$, $f=0.6$, and $f=0.7$, as
  shown. The thin vertical lines in these figures is the spot
  self-replication threshold $S_j=\Sigma_2(f)$ (see (\ref{self_rep})).
  For $S_j>\Sigma_2(f)$, the quasi-equilibrium spot solution is
  linearly unstable on an ${\mathcal O}(1)$ time-scale. In this figure, the
  values of $f$ decrease in the direction of the arrow.}
\label{UVchi}
\end{center}
\end{figure}

Given some value of $S_j$ and $f$, we solve (\ref{U0V0eqn})
numerically on the truncated domain $\rho \in [0, R]$, with $R \gg 1$,
where we impose the approximate conditions $U_{j0}(R) = 0$ and
$V_{j0}^{\prime}(R) = {S_j/R}$.  This yields solutions $U_{j0}$ and
$V_{j0}$, and we approximate $\chi$ by $\chi \approx V_{j0}(R)- S_j
\log R$. In Fig.~\ref{UV} we plot $U_{j0}$ for different values of
$S_j$ when $f = 0.3$ and $R=20$. In the left panel of Fig.~\ref{UVchi}
we plot $\chi$ versus $S_j$ for $f=0.3$. For $S_j\to 0$, the
asymptotic behavior of $\chi$, as derived in \cite{rozada2014}, is
\begin{equation}  \label{chi:small_S}
\begin{gathered}
   \chi(S_j) \sim \frac{d_0}{S_j} + d_1 S_j + \cdots\,, \quad \mbox{as} \quad
  S_j\to 0 \,,  \\
d_0 \equiv \frac{b(1-f)}{f^2}\,, \qquad d_1 = \frac{0.4893}{1-f} - 0.4698\,, 
\qquad
b \equiv \int_0^\infty \rho w^2 \, \de{\rho} \approx 4.934\,,
\end{gathered}
\end{equation}
where $w(\rho)>0$ is defined to be the unique solution of $\dell_\rho
w - w + w^2=0$ with $w\to 0$ as $\rho\to \infty$. In the right panel
of Fig.~\ref{UVchi} we plot $\chi$ versus $S_j$ for a few $f$ values.

\section{Proofs of Lemmas}\label{appendix:proofs}

\subsection{Proof to Lemma~\ref{lemma:coord} (Tangent plane approximation)}
\label{proof:coord}

We begin by letting $\bm{x}\equiv(\cos\phi \sin \theta, \sin\phi
\sin\theta, \cos\theta)^T \equiv (f_1,f_2,f_3)^T$, where
$f_i=f_i(\phi,\theta)$ for $i=1,2,3$. By retaining the quadratic terms
in the Taylor expansion of ${\bm x}$ as ${\bm x}\to {\bm x}_j$, we
readily derive that
\begin{subequations}\label{dmap}
\begin{equation}
   \bm{x}-\bm{x}_j \sim \eps \bm{J}_j {\bm s} + \frac{\eps^2}{2} {\bm r} +
 \cdots \,, \label{dmap_1}
\end{equation}
where $\bm{J}_j$ is defined in (\ref{map_3}) and $\bm{r}\equiv (r_1,r_2,r_3)^T$
with components defined by
\begin{equation}
   r_i \equiv \bm{s}^T {\mathcal H}_i \bm{s} \,, \qquad {\mathcal H}_i \equiv
   \begin{pmatrix}
     f_{i\theta\theta}  &  {f_{i\theta\phi}/\sin\theta} \\
     {f_{i\theta\phi}/\sin\theta}  &  {f_{i\phi\phi}/\sin^2\theta} \\
   \end{pmatrix}\Biggr\rvert_{\phi=\phi_j,\theta=\theta_j} \,, \qquad i=1,2,3\,.
  \label{dmap_2}
\end{equation}
The leading term in (\ref{dmap_1}) gives the first expression in
(\ref{map_1}).  To obtain the second relation in (\ref{map_1}), we
calculate $|\bm{x}-\bm{x}_j|^2 \sim \eps^2 \left( \bm{s}^T \bm{J}_j^T \bm{J}_j
\bm{s} + \eps \bm{s}^T \bm{J}_j^T \bm{r} \right)$.  Since
$\bm{J}_j^T \bm{J}_j=\I$ and $\bm{s}^T\bm{s}=s_1^2+s_2^2$, we obtain
\begin{equation}
   |\bm{x}-\bm{x}_j| \sim \eps \left(s_1^2+s_2^2\right)^{1/2} \left(
 1 + \frac{\eps}{2(s_1^2+s_2^2)} \bm{s}^T \bm{J}_j^{T} \bm{r}\right) \,.
\label{dmap_3}
\end{equation}
\end{subequations}
Finally, we use (\ref{map_3}) for $\bm{J}_j^T$ and we evaluate the
required partial derivatives in (\ref{dmap_2}) to calculate
$\bm{r}$. After some lengthy, but straightforward, algebra we get that
$\bm{s}^T \bm{J}_j^{T} \bm{r}= s_1 s_2^2 \cot\theta_j$. Upon
substituting this result into (\ref{dmap_3}) we obtain the second
result in (\ref{map_1}).

\subsection{Proof to Lemma~\ref{lemma:tangent} (Static component of first-order inner solution)}
\label{proof:tangent}
The proof is by a direct verification. We set
\begin{equation}\label{part:guess}
  \bm{U}_1=A s_2^2 \partial_{s_1} \bm{U}_0 + B s_1 s_2 \partial_{s_2}\bm{U}_0\,,
\end{equation}
for some constants $A$ and $B$. For this form of $\bm{U}_1$ we readily
calculate that
\begin{align*}
\dell_{(s_1,s_2)} \bm{U}_1 &= A s_2^2 \partial_{s_1}\left(\dell_{(s_1,s_2)} \bm{U}_0
 \right) + B s_1 s_2 \partial_{s_2} \left(\dell_{(s_1,s_2)} \bm{U}_0
\right) \\ &\qquad + s_2\left(4A+2B\right) \partial_{s_1s_2} \bm{U}_{0} 
   + 2 B s_1 \partial_{s_2s_2} \bm{U}_0 + 2 A \partial_{s_1} \bm{U}_0 \,.
\end{align*}
In this expression, we use
 $\partial_{s_1} \dell_{(s_1,s_2)} \bm{U}_0=-{\mathcal M} \partial_{s_1}\bm{U}_0$ 
and $\partial_{s_2} \dell_{(s_1,s_2)}\bm{U}_0=-{\mathcal M}\partial_{s_2}\bm{U}_0$, 
as obtained from differentiating (\ref{part_0}), to obtain
\begin{equation*}
\dell_{(s_1,s_2)} \bm{U}_1 = -A s_2^2 {\mathcal M} \partial_{s_1}
\bm{U}_0 - B s_1 s_2 {\mathcal M} \partial_{s_2} \bm{U}_0 +
s_2\left(4A+2B\right) \partial_{s_1s_2} \bm{U}_{0} + 2 B s_1
\partial_{s_2s_2} \bm{U}_0 + 2 A \partial_{s_1} \bm{U}_0 \,.
\end{equation*}
For $\bm{U}_1$ of the form (\ref{part:guess}) we then calculate that
${\mathcal M} \bm{U}_1 = A s_{2}^2 M \partial_{s_1} \bm{U}_0 + B s_1
s_2 {\mathcal M} \partial_{s_2} \bm{U}_0$. Upon adding these two
expressions, we obtain
\begin{equation*}
  \Lop \bm{U}_{1} \equiv \dell_{(s_1,s_2)} \bm{U}_{1} + {\mathcal M} \bm{U}_1 =
  2 s_2 \left(2A + B \right) \partial_{s_1s_2} \bm{U}_0 + 2 B s_1
  \partial_{s_2s_2} \bm{U}_0 + 2A \partial_{s_1} \bm{U}_0 \,.
\end{equation*}
The right hand-side of this expression agrees with that in (\ref{part_1a}) 
if we choose $2A=-\cot\theta_j$ and $B=\cot\theta_j$. Finally, we calculate
the far-field behavior of $V_{1}$ using (\ref{part:guess}). This yields
$V_{1} \sim S_j s_1 s_2^2 {(A+B)/\rho^2}= S_j s_1 s_2^2 {\cot\theta_j/(2\rho^2)}$
as $\rho\to \infty$, which agrees with (\ref{part_1b}).

\subsection{Proof to Lemma~\ref{lemma:bj} (Diagonal entries of $\B$)}
\label{proof:bj}
We shall derive (\ref{hatb1}) and establish (\ref{k:prop}). First,
note that as $S_j\to 0$, the solution to the core problem
(\ref{U0V0eqn}) is given by (Principal Result 4.1 of
\cite{rozada2014})
\begin{equation}
  U_{j0}\sim \frac{S_j w}{f \tilde{v}_0} \,, \qquad V_{j0}\sim \frac{\tilde{v}_0}
  {S_j} \,, \qquad \tilde{v}_0 \equiv \frac{b(1-f)}{f^2} \,, \label{core:smalls}
\end{equation}
where $w(\rho)>0$ is the unique solution to $\dell_\rho w - w + w^2=0$ with
$w(\infty)=0$, and $b\equiv \int_{0}^{\infty} \rho w^2\, \de{\rho}$. 
In (\ref{eqn:psi_N}), we then expand $\psi_j$, $N_j$ and $B_j$ for
$S_j\to 0$ as
\begin{equation}
 N_j= S_{j}^{-2} \left( \hat{N}_j + {\mathcal O}(S_j^2)\right) \,, \quad
 B_j= S_{j}^{-2} \left( \hat{B}_j + {\mathcal O}(S_j^2)\right) \,, \quad 
 \psi_j = \hat{\psi}_j + {\mathcal O}(S_j^2) \,. \label{eig:smalls}
\end{equation}
Upon substituting (\ref{core:smalls}) and (\ref{eig:smalls}) into
(\ref{eqn:psi_N}), and collecting powers of $S_j$, we obtain that
\begin{subequations}\label{eqn:tls5} 
\begin{gather}
  \dell_\rho \hat{\psi}_j - \hat{\psi}_j + 2w\hat{\psi}_j  -\lambda \hat{\psi}_j
  = -\frac{w^2}{f \tilde{v}_0^2}\hat{B}_j, \quad 
 \hat{\psi}_j^{\prime}(0)=0, \quad \hat{\psi}_j \to 0 
\text{ as $\rho \to \infty$}\,,  \label{eqn:tls5_a}\\
  \dell_{\rho}\hat{N}_j  = \hat{\psi}_j\left(\frac{2w}{f}-1\right) +
 \frac{w^2}{f^2\tilde{v}_0^2}\hat{B}_j, \quad
  \hat{N}_j^{\prime}(0)=0, \quad \hat{N}_j\sim\log\rho + {\mathcal O}(1)
  \text{ as $\rho\to\infty$}. 
\end{gather}
\end{subequations}
By integrating the equations for $\hat{N}_j$ and for
$\hat{\psi}_j$ over $0<\rho<\infty$, we obtain that
\begin{equation} \label{eqn:tls7}
 \frac{2}{f}\int_{0}^{\infty}\hat{\psi}_jw\rho \, \de{\rho} -
 \int_{0}^{\infty}\hat{\psi}_j\rho \, \de{\rho} + \frac{\hat{B}_j
 b}{f^2 \tilde{v}_0^2} = 1 \,, \qquad
 -(1+\lambda) \int_{0}^{\infty}\hat{\psi}_j\rho \, \de{\rho} +
2\int_{0}^{\infty}w\hat{\psi}_j\rho\, \de{\rho} = - 
\frac{\hat{B}_j b}{f\tilde{v}_0^2}\,.
\end{equation}
Upon eliminating $\int_{0}^{\infty} \hat{\psi}_j\rho \, \de{\rho}$
between these two expressions we obtain that
\begin{equation}
  \frac{1}{f} \int_{0}^{\infty} w \hat{\psi}_j \rho\, \de{\rho} +
 \frac{\hat{B}_j b}{2 f^2\tilde{v}_0^2} = \frac{\lambda+1}{2(\lambda+1 -f)}\,.
   \label{eqn:tmore}
\end{equation}
Then, in the class of radially symmetric solutions, we write the solution,
$\hat{\psi}_j$, to \eqref{eqn:tls5_a} as
\begin{equation}
 \hat{\psi}_j = -\frac{\hat{B}_j}{f \tilde{v}_0^2}\left( L_0 -
  \lambda\right)^{-1}w^2 \,, \qquad \mbox{where} \quad
  L_0 \Phi \equiv \dell_\rho \Phi - \Phi + 2 w\Phi \,. \label{eqn:tls6}
\end{equation}
Finally, upon substituting (\ref{eqn:tls6}) into (\ref{eqn:tmore}) and solving
for $\hat{B}_j$, we readily obtain \eqref{hatb1} of Lemma~\ref{lemma:bj}. 

Next, we establish (\ref{k:prop}) for $\K(\lambda)$ as defined in
(\ref{hatb1_2}). The self-adjoint problem $L_0\Phi=\sigma \Phi$ has a
unique real eigenvalue $\sigma_0>0$ with eigenfunction $\Phi_0>0$,
which we normalize as $\int_{0}^{\infty} \rho \Phi_0^2 \,
\de{\rho} = 1$. Since $L_0^{-1} w^2=w$, we get $\K(0)=b-{b/2}={b/2}$. The
monotonicity result $\K^{\prime}(\lambda)>0$ in (\ref{k:prop_1}) for
the segment $0<\lambda<\sigma_0$ of the real axis was proved in
Appendix C of \cite{rozada2014}.

To establish the asymptotics (\ref{k:prop_2}) as
$\lambda\to\sigma_0^{-}$, we introduce $\delta>0$ small and set
$\lambda=\sigma_0-\delta$. We then expand the solution $q$ to
$(L_0-\lambda) q=w^2$ as $q=C \delta^{-1}\Phi_0 + q_1+ \cdots$, for
some constant $C$ to be found. We obtain that $q_1$ satisfies
$(L_0-\sigma_0)q_1=w^2-C\Phi_0$, which has a solution only if
$C=\int_{0}^{\infty} \rho w^2 \Phi_0 \, \de{\rho}$. Thus, for
$\delta\ll 1$, we have $(L_0-\lambda)^{-1}w^2 \sim \delta^{-1} C
\Phi_0$. Upon substituting this expression into (\ref{hatb1_2}) we
obtain the asymptotics (\ref{k:prop_2}) when $\lambda
=\sigma_0-\delta$ with $\delta\ll 1$. Finally, to establish
(\ref{bj:p1}), we use $B_{j}(S_j,0)=\chi^{\prime}(S_j)$ at each $f>0$
and the asymptotics for $\chi(S_j)$ in (\ref{chi:small_S}) as $S_j\to
0$.

\subsection{Proof of Lemma~\ref{lemma:two-spot} (Explicit two-spot solution)}
\label{proof:two-spot}
For any two-spot configuration $\G$ satisfies (\ref{Ge}),
so that from (\ref{Sc}) we have $S_1=S_2=E$. This is the unique solution
to (\ref{constraint}) with $S_j={\mathcal O}(1)$ as $\nu\to 0$. Assume
that $E<\Sigma_2(f)$, so that the DAE dynamics (\ref{slow_cart}) is valid.
We use (\ref{slow_cart}) to calculate
\begin{equation*}
   \dd{|\bm{x}_j-\bm{x}_i|^2}{\sigma}= -2 \left( \bm{x}_2^T \bm{x}_1^{\prime}
   + \bm{x}_1^T \bm{x}_2^{\prime}\right) = -\frac{8E}{{\mathcal A}(E)
   |\bm{x}_2-\bm{x}_1|^2} \left( 1 - (\bm{x}_2^T \bm{x}_1)^2\right) \,.
\end{equation*}
Since $|\bm{x}_2-\bm{x}_1|^2=2(1-\cos\gamma_{1,2})$ and
$\bm{x}_2^T\bm{x}_1=\cos(\gamma_{1,2})$, the expression above reduces to
\begin{equation*}
2 \sin\gamma_{1,2} \, \dd{\gamma_{1,2}}{\sigma}= -\frac{4E}{{\mathcal A}(E)}
 \left( 1 + \cos\gamma_{1,2}\right) =-\frac{8E}{{\mathcal A}(E)}
   \cos^{2}\left({\gamma_{1,2}/2}\right) \,.
\end{equation*} 
Since ${\mathcal A}(E)<0$, this ODE is ${d\gamma_{1,2}/d\sigma}=2E
{\cot\left({\gamma_{1,2}/2}\right)/|{\mathcal A}(E)|}$, with solution
(\ref{2spot}).

\subsection{Proof of Lemma~\ref{lemma:refer_sphere} (Invariance under orthogonal transformations)}
\label{proof:refer_sphere}
The Green's matrix $\mathcal{G}$ in the
constraint (\ref{constraint}) is invariant under $\RC$ since $\RC^T
\RC=\I$ implies $|\bm{\xi}_j-\bm{\xi}_i|=|\bm{x}_j-\bm{x}_i|$ for
$i\neq j$. Then, multiply (\ref{slow_cart}) by $\RC$ and use
$\RC^T\RC=\I$ to get
\begin{equation*}
  \dd{\RC \bm{x}_j}{\sigma} = \frac{2}{\mathcal{A}_j} \left( \RC - 
\RC \bm{x}_j \bm{x}_j^T \RC^T \RC \right) \sum_{\substack{i=1 \\ i \neq j}}^N  
  \frac{S_i \bm{x}_i}{|\bm{x}_i-\bm{x}_j|^2}\,, \qquad
  \RC \bm{x}_j(0) = \RC \bm{x}_j^{0} \,, \quad j=1,\ldots, N\,.
\end{equation*}
The result follows by setting $\bm{\xi}_j=\RC \bm{x}_j$ 
and using $|\bm{\xi}_j-\bm{\xi}_i|=|\bm{x}_j-\bm{x}_i|$ for any $i\neq j$.

\section{Slow spot dynamics for the Schnakenberg model}\label{sec:other}

Results similar to those in Principal Results \ref{thm:quasi} and
\ref{thm:dyn} can be derived for other RD systems.  Here we focus on
the reduced Schnakenberg model formulated in
terms of a parameter $a>0$ as
\begin{equation} \label{s:full}
\pd{u}{t}  = \eps^2 \dell_S u -u + v u^2 \,, \qquad
\tau \pd{v}{t} = \dell_S v + a - \eps^{-2} u^2 v \,.
\end{equation}

\begin{figure}[htb]
\begin{center}
\includegraphics{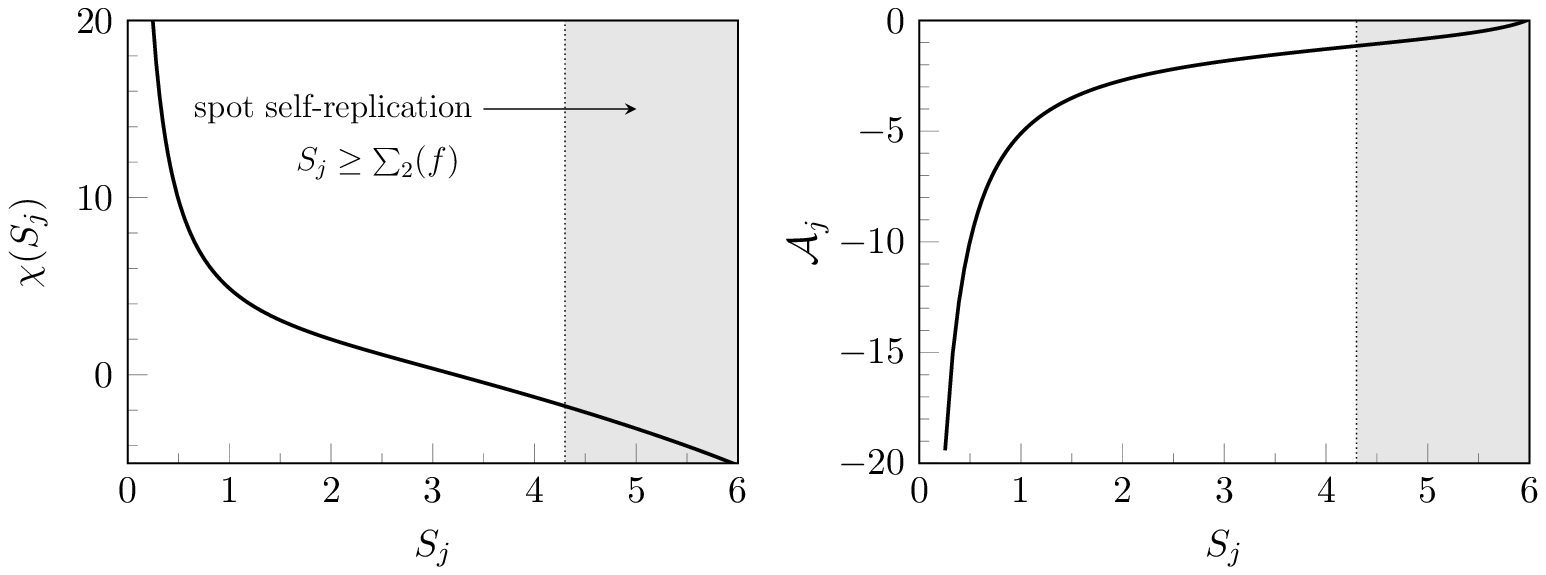}
\caption{Schnakenberg Model. Left: $\chi(S_j)$ versus $S_j$, computed
  from (\ref{s:U0V0eqn}). Right: ${\mathcal A}_j$ versus $S_j$. The
  shaded regions in these figures are the regions
  $S_j>\Sigma_2\approx 4.3$ where the spot is unstable on an 
  ${\mathcal O}(1)$ time-scale to a self-replication instability.
   \label{sfig:chi-adj}}
\end{center}
\end{figure}

In place of (\ref{U0V0eqn}), the leading-order radially symmetric
inner problem near the $j^\textrm{th}$ spot is given by solving, for
$0<\rho<\infty$, the coupled system
\begin{subequations}\label{s:U0V0eqn}
\begin{gather}
\dell_\rho U_{j0} - U_{j0} + U_{j0}^2 V_{j0} = 0 \,, \qquad \dell_\rho
V_{j0} - U_{j0}^2 V_{j0} = 0\,, \label{sU0V0eq} \\ 
U_{j0}^{\prime}(0) = V_{j0}^{\prime}(0)=0 \,;
\qquad U_{j0} \to 0 \,, \ V_{j0} \sim S_j \log \rho + \chi + o(1)
\,, \text{ as $\rho \to \infty$}\,. \label{sV0bc}
\end{gather}
\end{subequations}
The numerically computed function $\chi=\chi(S_j)$ is plotted in
the left panel of Fig.~\ref{sfig:chi-adj}.

The other function required for the slow dynamics, and which depends
on the specific form of the nonlinear kinetics, is ${\mathcal
  A}_j$ defined in (\ref{asolv}). In computing ${\mathcal A}_j$
from (\ref{asolv}), $U_{j0}$ is now given by the solution to
(\ref{s:U0V0eqn}) and $P_{1}(\rho)$ is the solution to (\ref{peq})
subject to $(P_1,P_2)^T\sim (0,{1/\rho})^T$ as $\rho\to \infty$, where
the matrix ${\mathcal M}_j$ in (\ref{peq}) is now given in terms of
the solution to (\ref{s:U0V0eqn}) by
\begin{equation}
    {\mathcal M}_j \equiv \begin{pmatrix}
                -1 + 2U_{j0} V_{j0} &  U_{j0}^2 \\
                 - 2U_{j0} V_{j0}   & -U_{j0}^2 \\
               \end{pmatrix}  \,.
\end{equation}
The computed function ${\mathcal A}_j$ versus $S_j$ for the
Schnakenberg model is plotted in Fig.~\ref{sfig:chi-adj}.  In terms of
these model-specific functions $\chi(S_j)$ and ${\mathcal A}_j$, the
result for slow spot dynamics is as follows:

\begin{result}[Schnakenberg model: slow spot dynamics] \label{thm:sdyn} 
Let $\eps\to 0$. Provided that there are no ${\mathcal O}(1)$
time-scale instabilities of the quasi-equilibrium spot pattern, the
slow dynamics of the spot pattern on the unit sphere for 
(\ref{s:full}) is characterized by the
quasi-equilibrium solution
\begin{equation}\label{squasi_eq}
 u_\unif \sim \sum_{i=1}^N U_{i,0}\left(\frac{|\bm{x}-\bm{x}_i|}{\eps}\right)\,, 
\qquad v_\unif \sim \sum_{i=1}^N S_i L_i(\bm{x}) +\frac{\overline{v}_c}{\nu}\,,
\end{equation}
where the time-dependent spot locations $\bm{x}_j(\sigma)$ on the slow
time-scale $\sigma$, with $\sigma = \eps^2 t$, satisfy 
\begin{subequations}\label{sslow_dyn}
\begin{equation}\label{sslow_cart}
  \dd{\bm{x}_j}{\sigma} = \frac{2}{\mathcal{A}_j} 
\left( \I - {\mathcal Q}_j\right) \sum_{\substack{i=1 \\ i \neq j}}^N  
\frac{S_i \bm{x}_i}{|\bm{x}_i-\bm{x}_j|^2}
 \,, \qquad {\mathcal Q}_j\equiv \bm{x}_j \bm{x}_j^T \,, 
\qquad j=1,\ldots,N \,,
\end{equation}
where $S_j$ for $j=1,\ldots,N$, and the constant $\overline{v}_c$ in
(\ref{squasi_eq}), are coupled to the spot locations and the parameter $a$
in (\ref{s:full}) by the $N$-dimensional nonlinear algebraic system
\begin{equation}\label{s:constraint}
\N(\bm{S}) \equiv \Bigl[ \I - \nu (\I - {\mathcal E}_0)\mathcal{G}\Bigr] 
\bm{S} + \nu (\I - {\mathcal E}_0) \bm{\chi}(\bm{S}) - \frac{2a}{N} 
\bm{e}=\bm{0} \,, 
\end{equation}
\end{subequations}
with $\overline{v}_c= 2a N^{-1} + \nu N^{-1} \left(\bm{e}^T
\bm{\chi}(\bm{S}) -\bm{e}^T {\mathcal G} \bm{S}\right)$.  In
(\ref{squasi_eq}) and (\ref{s:constraint}),
$L_i(\bm{x})\equiv\log|\bm{x}-\bm{x}_i|$, while the matrices
${\mathcal G}$, ${\mathcal E}_0$, and the vectors $\bm{\chi}$,
$\bm{e}$ are as defined previously in Principal Result
\ref{thm:quasi}.
\end{result}

In \cite{Kolokolnikov2009} it was shown that the $j^\textrm{th}$ spot is
linearly unstable on an ${\mathcal O}(1)$ time-scale to locally
non-radially symmetric perturbations near $\bm{x}_j$ when
$S_j>\Sigma_2\approx 4.3$. This linear instability was found in
\cite{Kolokolnikov2009} to lead to a nonlinear spot self-replication
event.  From Fig.~\ref{sfig:chi-adj}, we have ${\mathcal A}_j<0$ on
$0<S_j<\Sigma_2$, so that the slow dynamics of spots is repulsive.  We
emphasize that the DAE system (\ref{sslow_dyn}) is remarkably similar
in form to that for the Brusselator model in Principal Results 
\ref{thm:quasi}--\ref{thm:dyn}.

\end{appendix}


\end{document}